\documentclass[aps,pra,superscriptaddress,twocolumn]{revtex4-2}
\usepackage{xcolor}
\usepackage{natbib}
\usepackage{amsmath}
\usepackage{amssymb}
\usepackage{amsthm}
\usepackage{nicefrac}
\usepackage{graphicx}
\usepackage{bm}
\usepackage{url}
\usepackage{physics}

\newcommand{\vac}{\mathrm{vac}}
\newcommand{\bkg}{\mathrm{bg}}
\newcommand{\des}{\mathrm{des}}
\newcommand{\clk}{\mathrm{clk}}
\newcommand{\obj}{\mathrm{obj}}
\newcommand{\dobj}{\mathrm{d}} %dressed object

\newcommand{\im}[1]{\operatorname{Im}\left[#1\right]}
\newcommand{\re}[1]{\operatorname{Re}\left[#1\right]}

\begin{document}

\title{Limitations on bandwidth-integrated passive cloaking}

\author{Benjamin Strekha}
\affiliation{Department of Electrical and Computer Engineering, Princeton University, Princeton, New Jersey 08544, USA}

\author{Alessio Amaolo}
\affiliation{Department of Chemistry, Princeton University, Princeton, New Jersey 08544, USA}

\author{Jewel Mohajan}
\affiliation{Department of Electrical and Computer Engineering, Princeton University, Princeton, New Jersey 08544, USA}

\author{Pengning Chao}
\affiliation{Department of Mathematics, Massachusetts Institute of Technology, Cambridge, Massachusetts 02139, USA}

\author{Sean Molesky}
\affiliation{Department of Engineering Physics, Polytechnique Montréal, Montréal, Québec H3T 1J4, Canada}

\author{Alejandro W. Rodriguez}
\affiliation{Department of Electrical and Computer Engineering, Princeton University, Princeton, New Jersey 08544, USA}

\begin{abstract}
We present a general framework for the computation of structure-agnostic bounds on the performance of passive cloaks over a nonzero bandwidth.
We apply this framework in 2D to the canonical scenario of cloaking a circular object.
We find that perfect cloaking using a finite-sized isotropic cloak is impossible over any bandwidth, with the bounds scaling linearly with the bandwidth before saturating due to the finite size of the cloak and the presence of material loss.
The bounds also exhibit linear scaling with material loss in the cloak and linear scaling with the inverse of the radial thickness of the design region before saturation due to finite-size effects or the presence of material loss.
The formulation could readily find applications in the development of cloaking devices, setting expectations and benchmarks for optimal performance.
\end{abstract}

\maketitle

% \section{Introduction}

\textit{Introduction.---} The ability to engineer optical cloaking of objects to make them undetectable by observers has wide-ranging implications for radar~\cite{nicolai_fundamentals_2010}, obscurance~\cite{li_hiding_carpet_2008,ergin_3D_cloak_2010}, and sensing~\cite{alu_cloaking_sensor_2009} applications.
Several techniques have been proposed to achieve (near-)invisibility at a single operating optical wavelength, including transformation optics~\cite{pendry_controlling_2006}, scattering-cancellation via plasmonic~\cite{alu_plasmonic_2005} or mantle cloaking~\cite{alu_mantle_2009}, waveguide cloaking~\cite{tretyakov_broadband_2009}, transmission-line networks~\cite{alitalo_transmission_2008}, or the use of anomalous localized resonances~\cite{milton_cloaking_2006}, to name a few.
A fundamental and practically relevant question is whether one can achieve (near-)invisibility over a wide spectral window, e.g., the entire visible spectrum.
As pointed out by Pendry et al. using an argument based on the phase and group velocities of light in the presence of dispersion~\cite{pendry_controlling_2006}, and more recently by others using more formal arguments for the inevitable distortion of a pulse wave by materials with local response~\cite{miller_perfect_2006}, perfect cloaking of an isolated object in vacuum over a nonzero bandwidth is impossible due to causality; the finite speed of light and the presence of dispersion limits fully effective cloaking to a single frequency.
However, the Pendry et al. and Miller proofs say little about imperfect cloaking (a nonzero but small scattering cross section).
For instance, engineering a cloak to attain near-perfect invisibility at a single frequency does not preclude ``good'' cloaking over some nonzero bandwidth around that frequency.
Hashemi et al.~\cite{hashemi_diameter-bandwidth_2012} presented an alternative proof that agrees with the results of Pendry et al. and Miller for the impossibility of perfect cloaking over a nonzero bandwidth for physical, causal materials, but additionally derived scaling relations with implications for design, e.g., the allowed bandwidth over which one may expect practical cloaking scales inversely with the diameter of the cloaked object.

In this article, we adapt recently developed constrained optimization approaches~\cite{chao_physical_2022,molesky_global_2020,jelinek_fundamental_2021} to study the following question
related to bounds on cloaking performance: Given a specified cloak material along with an operating
frequency and bandwidth, what is the best possible performance of a passive cloaking device that can be formed out of that
material within a prescribed region surrounding a cloaked object?
In addition to capturing scaling behaviors, the presented formalism allows for quantitative assessments of expected performance of cloaks. Typical realizations of cloaks require anisotropic metamaterials with extreme and exotic properties, such as permittivities and permeabilities less than one and even equal to zero in some regions of the cloak~\cite{pendry_controlling_2006,cai_optical_cloaking_2007,schurig_calculation_2006}.
This raises the natural question of what is the best achievable performance using the simplest materials possible.
While the numerical bounds lack the transparency and intuition of analytic expressions, they can still provide insight into scaling properties and general trends, and even lead to fairly tight limits (within an order of magnitude of structures discovered via topology optimization).

\textit{Formulation.---}
In line with the majority of previous works on cloaking systems~\cite{hashemi_diameter-bandwidth_2012,monticone_invisibility_2016,cassier_bounds_2017,jelinek_fundamental_2021}, and as a proof of concept, we study the common scenario of a cloaked object illuminated by an incident plane wave emitted by a far-away source. We consider extinction, a nonnegative quantity for passive systems~\cite{jackson_classical_1999}, as the fundamental figure of merit, for the following reasons. First, it is better suited for describing the practical goals of cloaks as opposed to scattered or absorbed power alone: 
minimizing extinction minimizes the scattered field over both near- (absorbed power) and far-field  (scattered power) domains~\cite{jelinek_fundamental_2021}, implying that an object becomes truly undetectable regardless of the position of an observer as the figure of merit approaches zero.
Second, analyticity in the upper-half of the complex frequency plane means that one can exploit the residue theorem to evaluate the spectral average over a nonzero bandwidth via a single scattering calculation~\cite{kuang_computational_2020,hashemi_diameter-bandwidth_2012}, which greatly simplifies computations.

Throughout, we use the language of scattering theory laid out in previous descriptions of electromagnetic bounds~\cite{molesky_global_2020,molesky_fundamental_2020,venkataram_fundamental_2020-1}, wherein the $\mathbb{T}$ operator describing bound polarization currents in the medium is defined by the relation
\begin{equation}
    \mathbb{I} = \mathbb{T}\left(\mathbb{V}^{-1} - \mathbb{G}_{\bkg}\right) = \left(\mathbb{V}^{-1} - \mathbb{G}_{\bkg}\right)\mathbb{T}.
    \label{eq:clk_Tdefintion}
\end{equation} 
The operator $\mathbb{G}_{\bkg}$ represents the background Green's function, which for
vacuum satisfies $\left[ \nabla\times\nabla\times -
  \frac{\omega_{0}^{2}}{c^{2}}\right]\mathbb{G}_{\vac}(\mathbf{r},
\mathbf{r}'; \omega_{0}) = \frac{\omega_{0}^{2}}{c^{2}}\delta^{(3)}(\mathbf{r} - \mathbf{r}')$. (Note that in keeping with prior work, this definition includes an additional factor of $\frac{\omega_{0}^{2}}{c^{2}}$ compared to the standard convention~\cite{novotny_principles_2012}).
The $\mathbb{V}$ operator is the scattering potential (susceptibility) relative to this background medium (whatever additional material response was not included in the definition of $\mathbb{G}_{\bkg}$), and $\left|\mathbf{E}_{i}\right>$ and $\left|\mathbf{J}_{i}\right>$ are defined as the incident electric fields and electric currents in the background, respectively. 
When acting on incident fields, the $\mathbb{T}$ operator produces the generated current $|\mathbf{J}_{g}\rangle = -\frac{ik_{0}}{Z} \mathbb{T}|\mathbf{E}_{i}\rangle$,
with $Z = \sqrt{\mu_{0}/\epsilon_{0}}$ denoting the impedance of free space.

Employing the $\mathbb{T}$ operator relation to a scattering problem involving a prescribed incident field~\cite{molesky_global_2020}, the extinguished power may be written as,
\begin{equation}
    P_{\mathrm{ext}}(\omega_{0}) = 
    \frac{1}{2}\re{\left<\mathbf{E}_{i}|\mathbf{J}_{g}\right>} = 
    \frac{1}{2Z}\im{k_{0}\left<\mathbf{E}_{i}|\mathbb{T}|\mathbf{E}_{i}\right>},
    \label{eq:clk_eIncExt}
\end{equation}
where $k_{0}$ is left inside Im because it will be continued to the complex plane to map this quantity to a bandwidth-averaged quantity~\cite{kuang_computational_2020,hashemi_diameter-bandwidth_2012}. Namely, since real sources emit light over a nonzero bandwidth a key figure of merit is the bandwidth average of extinct power.
Considering a Lorentzian window function $L(\omega) \equiv \frac{\Delta\omega/\pi}{(\omega-\omega_0)^2 + \Delta\omega^2}$ centered at $\omega_0$ with a bandwidth $\Delta\omega$, the average extinction power for incident plane waves,
\begin{align}
    \langle P_{\mathrm{ext}} \rangle \equiv \int_{-\infty}^{\infty}P_{\mathrm{ext}}(\omega)L(\omega)\dd \omega \to P_{\mathrm{ext}}(\tilde{\omega}),
\end{align}
may be evaluated by closing the contour and picking up the residue at the pole of the Lorentzian in the upper-half complex frequency plane, yielding Eq.~\eqref{eq:clk_eIncExt} with all vectors and operators evaluated at a complex frequency $\tilde{\omega} \equiv \omega_{0} + i\Delta\omega$ and complex wave number $\tilde{k} \equiv (\omega_{0} + i\Delta\omega)/c$, simplifying the calculation so that one no longer needs to consider power at each individual frequency within the Lorentzian window.
Note also that working at a complex frequency requires the evaluation of $\chi(\tilde{\omega})$, which is mathematically equivalent to using modified materials at a real frequency $\omega_{0}$~\cite{hashemi_diameter-bandwidth_2012}.

To isolate the impact of the cloak on scattering from a fixed object, we let $\mathbb{G}_{\bkg}$ refer to the vacuum Green's function and shift the known scattering properties of the fixed object into the background of the scattering operator describing the cloak, with $\mathbb{G}_{\obj}$ denoting the Green's function of the fixed object in isolation, so that $\mathbb{G}_{\obj}$ satisfies $\left[ \nabla\times\nabla\times - \frac{\omega_{0}^{2}}{c^{2}}(\mathbb{V}_{\obj} + \mathbb{I})\right]\mathbb{G}_{\obj}(\mathbf{r},
\mathbf{r}';
\omega_{0}) = \frac{\omega_{0}^{2}}{c^{2}}\mathbb{I}\delta^{(3)}(\mathbf{r} - \mathbf{r}')$.
In particular, let $\mathbb{T}_{\obj} = (\mathbb{V}_{\obj}^{-1} - \mathbb{G}_{\vac})^{-1}$ be the scattering operator of the fixed object (the object to be cloaked) in isolation, and let $\mathbb{T}_{\dobj} = (\mathbb{V}_{\clk}^{-1} - \mathbb{G}_{\obj})^{-1}$ be the scattering operator of the designable object (the cloak) dressed by the fixed object. After some straightforward algebraic manipulations (see Appendix), we find
\begin{multline}
    P_{\mathrm{ext}}(\omega_{0})
    = \frac{1}{2Z}\im{k_{0}\left<\mathbf{E}_{i}\right|\mathbb{G}_{\vac}^{-1}\mathbb{G}_{\obj}\mathbb{V}_{\obj}\left|\mathbf{E}_{i}\right>}  \\
    + \frac{1}{2Z}\im{k_{0}\left<\mathbf{E}_{i}\right|\mathbb{G}_{\vac}^{-1}\mathbb{G}_{\obj}\mathbb{T}_{\dobj}(\mathbb{G}_{\vac}\mathbb{T}_{\obj} + \mathbb{I})\left|\mathbf{E}_{i}\right>}.
\end{multline}
% \begin{align}
%     P^{\mathrm{ext}}_{\mathrm{flx}}(\omega_{0})
%     &= \frac{1}{2Z}\im{k_{0}\left<\mathbf{E}_{i}\right|\mathbb{G}_{\vac}^{-1}\mathbb{G}_{\obj}\mathbb{V}_{\obj}\left|\mathbf{E}_{i}\right>} \nonumber \\
%     &+ \frac{1}{2Z}\im{k_{0}\left<\mathbf{E}_{i}\right|\mathbb{G}_{\vac}^{-1}\mathbb{G}_{\obj}\mathbb{T}_{\dobj}(\mathbb{G}_{\vac}\mathbb{T}_{\obj} + \mathbb{I})\left|\mathbf{E}_{i}\right>}.
% \end{align}
The first term is the extinguished power when no cloak is present while only the second term depends on the cloaked object and quantifies the interaction between the cloak and cloaked object.

Our derivation of bounds exploits the optimization procedure based on Lagrange duality laid out in Refs.~\cite{boyd_convex_2004,chao_physical_2022}.
The loosest such bound only imposes that the optimal scattering operator satisfies the
conservation of power (optical theorem~\cite{jackson_classical_1999})
over the entire design domain, and not the full scattering equations.
Defining $|\mathbf{E}_{\dobj}\rangle \equiv (\mathbb{G}_{\vac}\mathbb{T}_{\obj} + \mathbb{I})|\mathbf{E}_{i}\rangle = \frac{iZ}{k_{0}}\mathbb{G}_{\obj}\left|\mathbf{J}_{i}\right>$, the total field in the presence of only the fixed object, and $ |\mathbf{T}_{\dobj}\rangle \equiv \mathbb{T}_{\dobj}|\mathbf{E}_{\dobj}\rangle$ so that $-\frac{ik_{0}}{Z}|\mathbf{T}_{\dobj}\rangle$ is the induced current in the cloak, we relax the problem such that, instead of optimizing over $\mathbb{T}_{\dobj}$ with support in the design region, we optimize over $|\mathbf{T}_{\dobj}\rangle$ with support in the design region, i.e., optimize over all possible polarization currents within the design region so that the bound considered here automatically takes into account all possible distributions of vacuum and the prescribed cloak material within the design region.
It should be noted, however, that this bound formalism does not yield the optimal material distribution of the cloak~\cite{chao_physical_2022}, a task that calls for more computationally demanding topology optimization methods (NP-hard problems that preclude guarantees of globally optimal solutions)~\cite{molesky_inverse_2018}.
Similar techniques have recently been used to derive bounds on deterministic scattering and fluctuational electrodynamic phenomena~\cite{molesky_global_2020,molesky_hierarchical_2020,venkataram_fundamental_2020-1,chao_maximum_2022,angeris_computational_2019,shim_fundamental_2019,amaolo_can_heterostructures_2023,mohajan_fundamental_2023,strekha_suppressing_2024,amaolo_raman_2024,venkataram_fundamental_2020-2,strekha_tracenoneq_2022,strekha_traceeq_2024}.
Concretely, we wish to solve the following problem:
\begin{subequations}
\begin{align}
\underset{\left|\textbf{T}_{\dobj}\right>}{\text{min}} 
\,\,\frac{1}{2Z}&\im{k_{0}\left<\mathbf{E}_{i}\right|\mathbb{G}_{\vac}^{-1}\mathbb{G}_{\obj}\left(\mathbb{V}_{\obj}\left|\mathbf{E}_{i}\right>+\mathbb{P}_{\des}\left|\textbf{T}_{\dobj}\right>\right)} 
    \label{eq:cloakingPrimalProblem}
    \\
\text{s.t.}\, \forall m, 
\quad  &\langle \mathbf{E}_{\dobj}|\mathbb{P}_{m}|\mathbf{T}_{\dobj}\rangle
    -\langle \mathbf{T}_{\dobj}|\mathbb{U}_{m}|\mathbf{T}_{\dobj}\rangle = 0,
%    \\
 %   \text{Re}[&\langle \mathbf{E}_{\dobj}|\mathbb{P}_{m}|\mathbf{T}_{\dobj}\rangle
 %   -\langle \mathbf{T}_{\dobj}|\mathbb{U}_{m}|\mathbf{T}_{\dobj}\rangle] = 0,
\end{align}
\end{subequations}
where $\mathbb{P}_{\des}$ is a projection operator into the design region $\Omega_{\des}$, $\mathbb{P}_{m}$ is a projection operator into a subregion $\Omega_{m} \subseteq \Omega_{\des}$ contained within the design region, and $\mathbb{U}_{m} \equiv \mathbb{P}_{\des}(\chi_{\clk}^{-1\dagger}\mathbb{P}_{m} - \mathbb{G}_{\obj}^{\dagger}\mathbb{P}_{m})\mathbb{P}_{\des}$.
Taking the real and imaginary parts of the complex-valued constraints enforce the conservation of resistive and reactive power, respectively, spatially integrated over a region $\Omega_{m}\subseteq \Omega_{\des}$~\cite{chao_physical_2022}.
This is a quadratically constrained quadratic program over the field $|\mathbf{T}_{\dobj}\rangle$ and we numerically compute bounds on Eq.~\eqref{eq:cloakingPrimalProblem}, the primal objective, by evaluating and optimizing the corresponding concave Lagrange dual function~\cite{boyd_convex_2004}.

\begin{figure*}
    \centering
        \includegraphics[width=\linewidth]{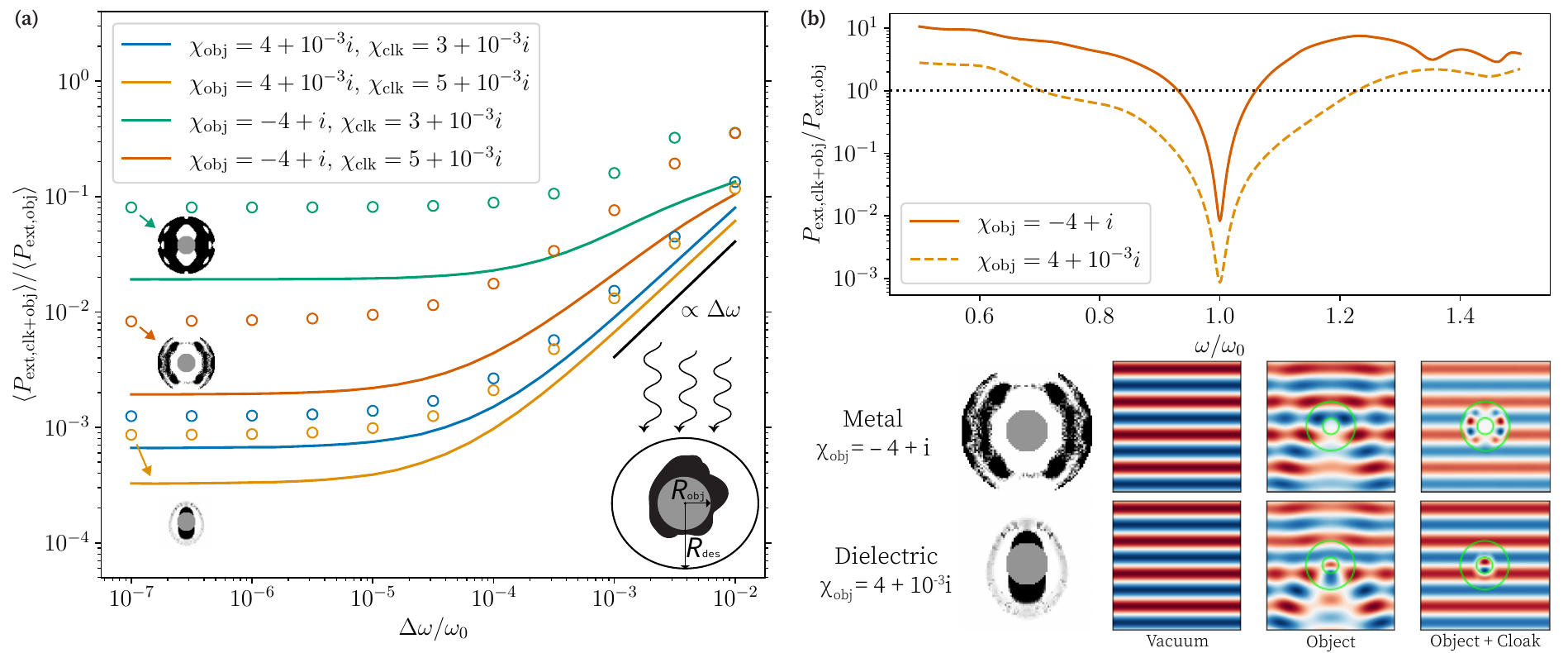}
        \caption{
        (a) Bounds on bandwidth-integrated extinction power as a function of the bandwidth.
        All curves and markers refer to performance bounds and inverse designs, respectively, expressed as a ratio of the extinguished power from incident plane waves in the presence and absence of a cloak for $R_{\obj} = \lambda_{0}/4$ and $R_{\des} = 3\lambda_{0}/4$ (bottom-right schematic).
        Insets: Representative inverse designs corresponding to $\Delta\omega/\omega_{0} = 10^{-7}$.
        (b) Performance of structures discovered via topology optimization for $\Delta\omega/\omega_{0}  = 10^{-7}$, $R_{\obj} = \lambda_{0}/4$, $R_{\des} = 3\lambda_{0}/4$, $\chi_{\clk}(\tilde{\omega}) = 5 +10^{-3}i$, and $\chi_{\obj}(\tilde{\omega}) = -4 + i$ (solid) or $\chi_{\obj}(\tilde{\omega}) = 4 + 10^{-3}i$ (dashed).
        Top: The cloak suppresses extinction within a frequency range around $\omega_{0}$ at the expense of worse performance in other parts of the spectrum.
        Bottom: Real part of the total out-of-plane electric field at $\omega_{0}$. Green circles outline the region of space where the object and cloak are contained, with an inner and outer radius $R_{\obj}$ and $R_{\des}$, respectively.
        }
\label{fig:cloak_bounds_vs_bandwidth}
\end{figure*}

% \section{Applications}

\textit{Applications.---} We now exploit the framework above to numerically obtain performance bounds for passive cloaks composed of an isotropic electric susceptibility. For simplicity, we focus on 2D settings and consider the scenario of a fixed circular object of radius $R_{\obj}$, a common benchmark example for cloaking~\cite{fleury_invisibility_2015}, with the cloak restricted to an annular design region with an inner radius $R_{\obj}$ and outer radius $R_{\des}$ [see Fig.~\ref{fig:cloak_bounds_vs_bandwidth}(a) bottom-right schematic] for the case of incident TM (electric field out of plane) plane waves.

First, we consider the situation of illumination of broadband light from the far-field (plane waves). For a fixed $\chi(\omega_{0} + i\Delta\omega)$ we find that the bounds scale linearly with the bandwidth before saturating due to the finite size of the design domain and due to the presence of material loss, see Fig.~\ref{fig:cloak_bounds_vs_bandwidth}(a).
As seen in the figure, enforcing the conservation of power within the design domain leads to nontrivial bounds with interesting trends and scaling behavior that are also fairly tight (within an order of magnitude of the objective values of discovered structures).
The structures found via topology optimization perform well within a bandwidth $\Delta\omega$ around the ``center'' frequency $\omega_{0}$, but at the expense of more scattering in nearby regions of the spectrum, see Fig.~\ref{fig:cloak_bounds_vs_bandwidth}(b).
This is in agreement with the findings in Ref.~\cite{monticone_cloaked_2013}, which showed that any passive cloak made of a linear, nondiamagnetic material respecting causality always increases the scattering and extinction integrated over all frequencies compared to the original uncloaked object.
Thus, near-invisibility in a given frequency window necessarily implies significant scattered/extinct power at other frequencies, and therefore such cloaks may be detected more easily than the original uncloaked objects when illuminated by sufficiently broadband light (short pulses).
As seen in Fig.~\ref{fig:cloak_bounds_vs_bandwidth}(b), the electric field in the lossy metallic object is expelled from and rerouted around the object with the introduction of the cloak. The dielectric object with a smaller loss, on the other hand, 
may try to create zero-field regions or it may allow the electric field to penetrate the object if it is advantageous as is the case for the dielectric cloak example shown in Fig.~\ref{fig:cloak_bounds_vs_bandwidth}(b).

\begin{figure}
    \centering
        \includegraphics[width=\linewidth]{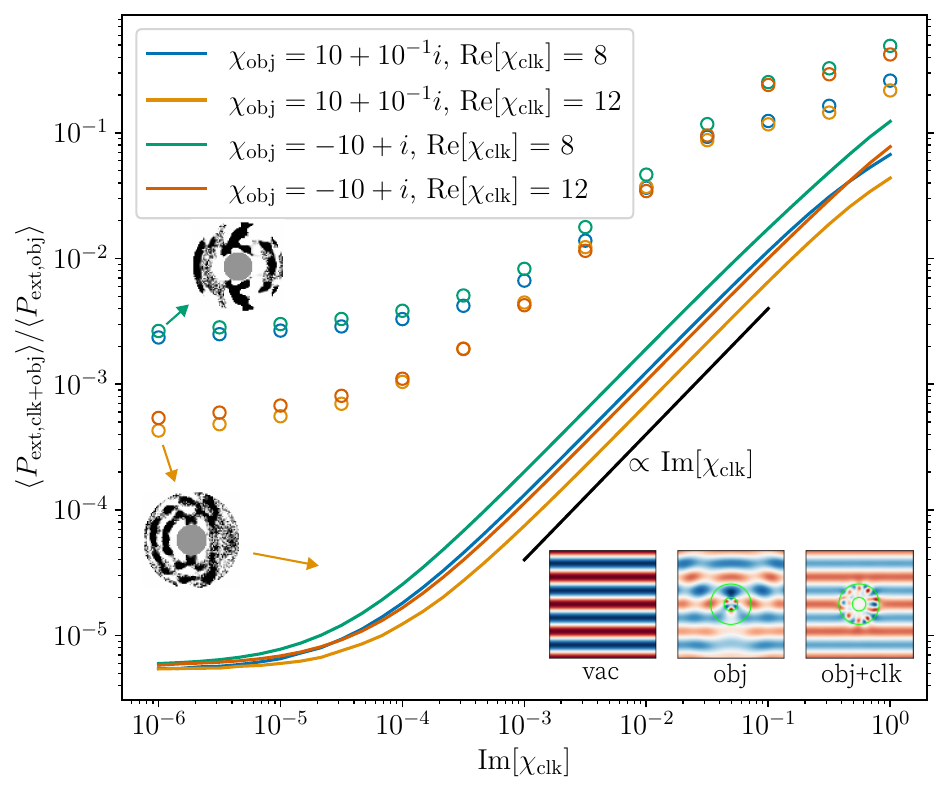}
        \caption{
        Bounds (solid lines) and inverse designs (circles) relating to net extinction power (cloaking performance) for the system of Fig.~1, at a single frequency $\omega_{0} = 2\pi c/\lambda_{0}$ and as a function of the material loss in the cloak for $R_{\obj} = \lambda_{0}/4$ and $R_{\des} = 3\lambda_{0}/4$, and for both dielectric and metallic objects.
  %      Topology optimization with random initialization discovers better-performing devices compared to vacuum initialization (not shown) but leads to asymmetry of the cloak with respect to the vertical axis. % (the direction of the incident plane wave).
        }
    \label{fig:bounds_vs_imchid}
\end{figure}

Next, we investigate the performance limits of 2D cloaks of isotropic susceptibility as a function of the material loss, see Fig.~\ref{fig:bounds_vs_imchid}.
The bounds (solid lines) follow trends seen in topology-optimized designs (circular markers) over a broad range of material loss values and both demonstrate, unsurprisingly, improved performance as the material loss in the cloak decreases, exhibiting linear scaling with decreasing material loss before eventually saturating. %due to the finite footprint of the cloak.
This scaling can be understood as follows.
Writing the total $\mathbb{T}$ operator in $2\times 2$ block-form over the two regions, namely,
\begin{align}
    \mathbb{G}^{\vac} = 
    \begin{bmatrix}
        \mathbb{G}^{\vac}_{\obj,\obj} & \mathbb{G}^{\vac}_{\obj,\clk} \\
        \mathbb{G}^{\vac}_{\clk,\obj} & \mathbb{G}^{\vac}_{\clk,\clk}
    \end{bmatrix},
\end{align}
\begin{align}
    \mathbb{V}^{-1} =
    \begin{bmatrix}
        \mathbb{V}^{-1}_{\obj} & 0 \\
        0 & \mathbb{V}^{-1}_{\clk}
    \end{bmatrix},
\end{align}
\begin{align}
        \mathbb{T}^{-1} =
    \begin{bmatrix}
        \mathbb{T}_{\obj}^{-1} & -\mathbb{G}^{\vac}_{\obj,\clk} \\
        -\mathbb{G}^{\vac}_{\clk,\obj} & \mathbb{T}_{\clk}^{-1}
    \end{bmatrix},
\end{align}
leads, after the use of the Woodbury formula for a matrix inverse, to a total $\mathbb{T}$ operator expression of the form
\begin{align}
    \mathbb{T} 
    % &=
    % \begin{bmatrix}
    %     (\mathbb{T}_{\obj}^{-1} - \mathbb{G}^{\vac}_{\obj,\clk}\mathbb{T}_{\clk}\mathbb{G}^{\vac}_{\clk,\obj})^{-1} & \mathbb{T}_{\obj}\mathbb{G}^{\vac}_{\obj,\clk}(\mathbb{T}_{\clk}^{-1} - \mathbb{G}^{\vac}_{\clk,\obj}\mathbb{T}_{\obj}\mathbb{G}^{\vac}_{\obj,\clk})^{-1} \\
    %     \mathbb{T}_{\clk}\mathbb{G}^{\vac}_{\clk,\obj}(\mathbb{T}_{\obj}^{-1} - \mathbb{G}^{\vac}_{\obj,\clk}\mathbb{T}_{\clk}\mathbb{G}^{\vac}_{\clk,\obj})^{-1} & (\mathbb{T}_{\clk}^{-1} - \mathbb{G}^{\vac}_{\clk,\obj}\mathbb{T}_{\obj}\mathbb{G}^{\vac}_{\obj,\clk})^{-1}
    % \end{bmatrix} \\
    &\equiv 
    \begin{bmatrix}
        \mathbb{Y}_{\obj} & \mathbb{T}_{\obj}\mathbb{G}^{\vac}_{\obj,\clk}\mathbb{Y}_{\clk} \\
        \mathbb{T}_{\clk}\mathbb{G}^{\vac}_{\clk,\obj}\mathbb{Y}_{\obj} & \mathbb{Y}_{\clk}
    \end{bmatrix}.
    \label{eq:clk_2by2Toperatorform}
\end{align}
Here, we introduced $\mathbb{Y}_{\clk}$ which is the scattering operator of the designable cloak dressed by the presence of the fixed object, as seen from $\mathbb{Y}_{\clk} \equiv (\mathbb{T}_{\clk}^{-1} - \mathbb{G}^{\vac}_{\clk,\obj}\mathbb{T}_{\obj}\mathbb{G}^{\vac}_{\obj,\clk})^{-1} = (\mathbb{V}_{\clk}^{-1} - (\mathbb{G}^{\vac}_{\clk,\clk} + \mathbb{G}^{\vac}_{\clk,\obj}\mathbb{T}_{\obj}\mathbb{G}^{\vac}_{\obj,\clk}))^{-1}$ and recognizing $\mathbb{G}_{\vac} +\mathbb{G}_{\vac}\mathbb{T}_{\obj}\mathbb{G}_{\vac} = \mathbb{G}_{\obj}$ as the Green's function of the fixed object in isolation in vacuum.
Likewise for $\mathbb{Y}_{\obj}$ but with the fixed and design regions switching roles.
Letting $\chi_{\clk} = \chi_{\clk}' + i\chi_{\clk}''$ be the separation of the susceptibility into its real and imaginary parts, we find
\begin{multline}
    \mathbb{T}_{\clk}
    = 
    (\chi_{\clk}'\mathbb{P}_{\clk})[(\mathbb{I} - \mathbb{G}^{\vac}_{\clk,\clk}\mathbb{V}_{\clk}')^{-1} \\
    + \sum_{n=0}^{\infty}n(\mathbb{G}^{\vac}_{\clk,\clk})^{n}(\chi_{\clk}')^{n-1}(i\chi_{\clk}'')] \\ + (i\chi_{\clk}''\mathbb{P}_{\clk})(\mathbb{I} - \mathbb{G}^{\vac}_{\clk,\clk}\mathbb{V}_{\clk}')^{-1} + O[(\chi_{\clk}'')^{2}]
\end{multline}
and similar for $\mathbb{Y}_{\clk}$ but with $\mathbb{G}_{\vac}$ replaced with $\mathbb{G}_{\obj}$.
Using these expansions for $\mathbb{T}_{\clk}$ and $\mathbb{Y}_{\clk}$ in the $2\times 2$ block-form for the total scattering operator, Eq.~\eqref{eq:clk_2by2Toperatorform}, leads to the prediction that the extinction at a single frequency, $P_{\mathrm{ext}}(\omega) = \frac{k_{0}}{2Z}\left<\mathbf{E}_{i}|\im{\mathbb{T}}|\mathbf{E}_{i}\right>$, scales linearly with $\chi_{\clk}''$ towards a constant for a fixed geometric structure as $\chi_{\clk}''\to 0$.
The numerical Lagrange dual bounds follow the same scaling behavior as the primal objective function for a fixed structure, demonstrating that faster scaling is not possible.
The approach to a positive constant rather than 0 as the material loss of the cloak vanishes is in agreement with the findings of Hashemi et al.~\cite{hashemi_general_2011},
which showed that if the attainable refractive index contrast is bounded from below (as is the case for a cloak of isotropic susceptibility considered here) then there is a bound on the reduction of the scattering cross section for transformation-based invisibility cloaking of an isolated object; in particular, a minimum achievable refractive index contrast greater than 0 necessarily implies a positive cross section.
A Pendry cloak can achieve perfect invisibility at a single frequency with a finite device footprint but it requires a vanishing permittivity and permeability at the inner surface of the cloak~\cite{pendry_controlling_2006,cai_optical_cloaking_2007}.
As shown, however, the presented formalism can provide quantitative lower bounds supporting such qualitative observations.

\begin{figure}
    \centering
        \includegraphics[width=\linewidth]{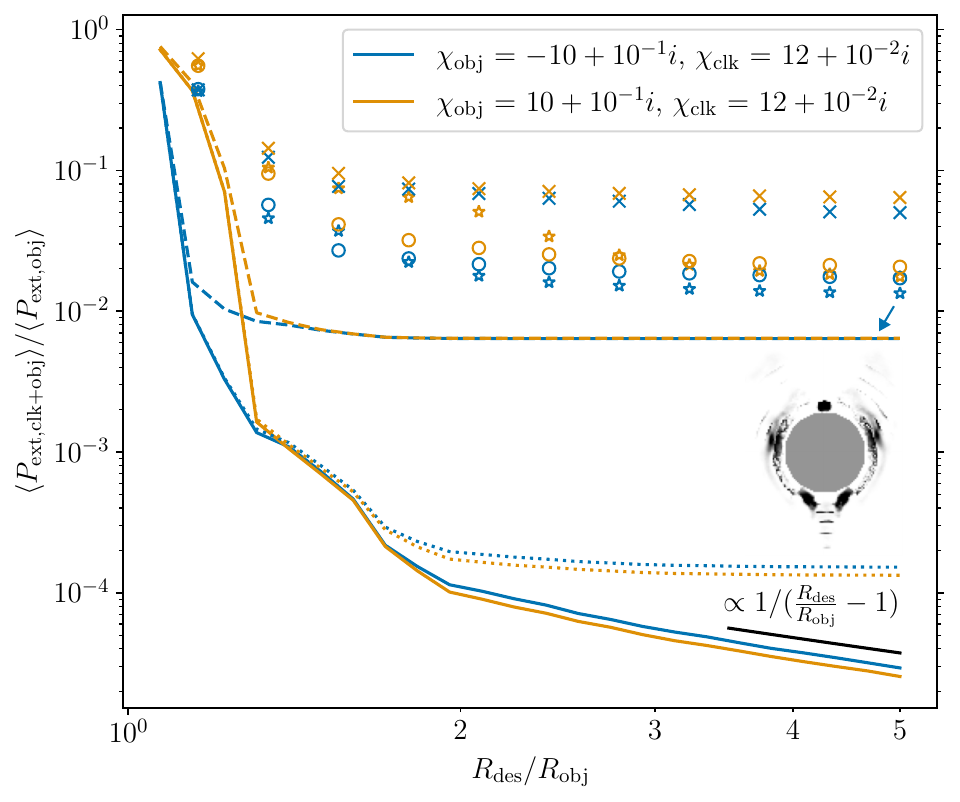}
        \caption{
        Bounds on extinction power for the same system as Fig.~1 but as a function of the outer radius of the design region $R_{\des}/R_{\obj}$, for a fixed object radius $R_{\obj} = \lambda_{0}/2$, and for different values of $\Delta\omega/\omega_{0} = 10^{-3}$ (dashed), $10^{-5}$ (dotted),  and zero bandwidth (solid); for comparison, crosses, circles and stars, respectively, are performance values obtained from inverse designs.  }
\label{fig:cloak_bounds_Rinner0d5_Router}
\end{figure}

Figure~\ref{fig:cloak_bounds_Rinner0d5_Router} plots limit values as a function of the radial size of the design region, for a given object radius.
Unsurprisingly, the lower bounds are monotonically nonincreasing as the allowed cloak footprint increases since any polarization current with a given performance can be contained in an enlarged design domain. 
In the single frequency case, $\Delta\omega \to 0$, the bounds scale inversely with the thickness of the design region, $R_{\des} - R_{\obj}$, over a notable range of thicknesses.
Hashemi et al.~\cite{hashemi_general_2011} argued that for bounded refractive indices, the cloak thickness must scale proportionally to the thickness of the object being cloaked, and the scaling of our numerical bounds demonstrates this behavior.
For positive bandwidths, $\Delta\omega > 0$, the bounds saturate to a positive value, with saturation occurring rather quickly (around $R_{\des}/R_{\obj} \approx 2$) for bandwidths $\Delta\omega/\omega_{0} \gtrsim 10^{-5}$.
This has interesting implications for the design of cloaks since it is typically not clear how large a device needs to be to achieve near-optimal or even reasonable performance.
For our chosen parameters, the bounds show that for nonzero bandwidths a cloak design region with a thickness equal to that of the cloaked object essentially saturates the possible performance; larger device footprints will not always lead to significant improvements in performance.

\begin{figure}
    \centering
        \includegraphics[width=\linewidth]{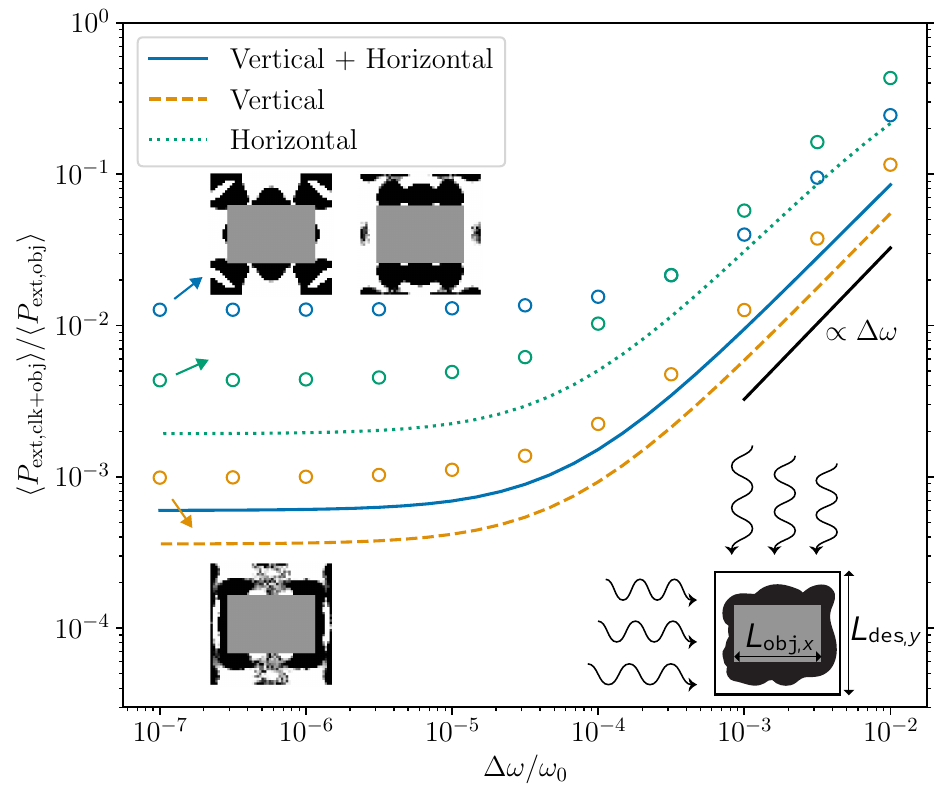}
        \caption{
        Bounds on bandwidth-integrated extinction as a function of the bandwidth with either vertically or horizontally incident plane waves, or an objective function with an equally weighted average of both (bottom-right schematic).
        The rectangular object has lengths $L_{\obj,x} = 3\lambda_{0}/4$ and $L_{\obj,y} = \lambda_{0}/2$ while the rectangular design region has lengths $L_{\des,x} = L_{\des,y} = \lambda_{0}$.
        The susceptibility values are $\chi_{\obj}(\tilde{\omega}) = 4 + 10^{-3}i$ and $\chi_{\clk}(\tilde{\omega}) = 5 + 10^{-3}i$.
        }
\label{fig:bounds_vs_bandwidth_2sides_rect}
\end{figure}

Lastly, the presented inverse designs and bounds were restricted to plane-wave sources coming from a single direction of incidence.
To consider cloaking robust to the direction of incidence, the figure of merit may be modified to encompass multiple plane wave directions (including fully angle-integrated extinction), such that $\langle P_{\mathrm{ext}}\rangle = 
  \sum_{a = 1}^{N}\frac{1}{2Z}\im{\tilde{k}\left<\mathbf{E}_{i}^{(a)}|\mathbb{T}|\mathbf{E}_{i}^{(a)}\right>}$
where each $|\mathbf{E}_{i}^{(a)}\rangle$ is a plane wave incident from a different angle.
This leads to an optimization problem of the form
\begin{subequations}
\begin{align}
\underset{\{\left|\textbf{T}_{\dobj}^{(a)}\right>\}}{\text{min}} 
\,\,
\sum_{a=1}^{N}\frac{1}{2Z}&\im{\tilde{k}\left<\mathbf{E}_{i}^{(a)}\right|\mathbb{G}_{\vac}^{-1}\mathbb{G}_{\obj}\mathbb{V}_{\obj}\left|\mathbf{E}_{i}^{(a)}\right>} \nonumber \\
+
\sum_{a=1}^{N}\frac{1}{2Z}&\im{\tilde{k}\left<\mathbf{E}_{i}^{(a)}\right|\mathbb{G}_{\vac}^{-1}\mathbb{G}_{\obj}
\mathbb{P}_{\des}\left|\textbf{T}_{\dobj}^{(a)}\right>} 
    \label{eq:cloakingPrimalProblemManyAngles}
    \\
\text{s.t.}\, \forall a, b, m, 
\quad  &\langle \mathbf{E}_{\dobj}^{(a)}|\mathbb{P}_{m}|\mathbf{T}_{\dobj}^{(b)}\rangle
    -\langle \mathbf{T}_{\dobj}^{(a)}|\mathbb{U}_{m}|\mathbf{T}_{\dobj}^{(b)}\rangle = 0,
\end{align}
\end{subequations}
where $|\mathbf{E}_{\dobj}^{(a)}\rangle \equiv (\mathbb{G}_{\vac}\mathbb{T}_{\obj} + \mathbb{I})|\mathbf{E}_{i}^{(a)}\rangle$ and $ |\mathbf{T}_{\dobj}^{(a)}\rangle \equiv \mathbb{T}_{\dobj}|\mathbf{E}_{\dobj}^{(a)}\rangle$.
Without the constraints between the $a\neq b$ terms, the problem reduces to a sum of decoupled optimization problems. 
Although each angle is a different scattering problem, 
the polarization currents induced in each scenario are generated by the same structured media and this fact is enforced by the $a\neq b$ ``cross constraints'', leading to additional tightening of the bounds~\cite{molesky_mathbbt-operator_2021,shim_multi-functional_2024}.
Since the incident field is the same up to a rotation, if the object and design region are invariant under rotations then the bounds are the same for decoupled problems and any tightening should come from the cross-constraints. 
To consider a more complicated scenario involving more than a single incident source, Fig.~\ref{fig:bounds_vs_bandwidth_2sides_rect}
shows computed bounds and inverse design performance values for a rectangular object discovered starting from random initializations for the case where $N = 2$, comprising horizontally and vertically propagating plane waves. 

% \begin{figure}
%     \centering
%         \includegraphics[width=\linewidth]{cloaking_inv_designs_bounds_2sides_FULL.pdf}
%         \caption{
%         Bounds on bandwidth-integrated extinction as a function of bandwidth, with the same parameters as in Fig.~\ref{fig:cloak_bounds_vs_bandwidth} but where the pluses ($+$) correspond to inverse designs for a figure of merit with a weighted-sum of incident vertical and horizontal plane waves (bottom-right schematic).
%         }
% \label{fig:bounds_vs_bandwidth_2sides}
% \end{figure}

% \section{Conclusion}

\textit{Conclusion.---} 
In conclusion, we presented a formalism for computing bounds on passive, broadband cloaking systems that can provide benchmarks and set expectations for future cloaking devices, particularly those developed through the use of topology optimization.
The many variables available for tuning means an exhaustive study of their relationship to the bounds on performance is complex.
We presented a few key examples to demonstrate the capabilities of the formalism and to extract intuition for the obvious parameters one may wish to vary in experiments.
Our numerical results support the claims that broadband cloaking of electrically large objects is practically impossible using passive linear cloaks~\cite{hashemi_diameter-bandwidth_2012,fleury_invisibility_2015,monticone_invisibility_2016} and provide a quantitative assessment of the limits on performance.
We believe this represents an important result for the science of cloaking, suggesting that new concepts and designs, including opening to the field of nonlinear and active metamaterials~\cite{fleury_invisibility_2015}, are necessary for increasing the cloaking bandwidth
and compensating the unwanted scattering required by linearity and causality in the presence of absorption losses~\cite{monticone_invisibility_2016}.
In closing, we note that while our examples focused on 2D and isotropic permittivities, the formalism is valid in 3D settings, and may be easily extended to consider anisotropic permittivities and permeabilities.

% \begin{acknowledgments}
\textit{Acknowledgments.---} 
We acknowledge the support of a Princeton SEAS Innovation Grant and by the Cornell Center for Materials Research (MRSEC) through Award DMR-1719875.
S.M. acknowledges financial support from the Canada First Research Excellence Fund via the Institut de Valorisation des Données (IVADO) collaboration.
The simulations presented in this article were performed on computational
resources managed and supported by Princeton Research
Computing, a consortium of groups including the Princeton Institute for Computational Science and Engineering
(PICSciE) and the Office of Information Technology's
High Performance Computing Center and Visualization Laboratory at Princeton University.
% The views, opinions, and findings expressed herein are those of the authors and should not be interpreted as representing the official views or policies of any institution. 
% \end{acknowledgments}

\appendix

\section{Objective function derivation}
\label{sec:app_obj_fun}

In this section, we derive an expression for the total (including the fixed and designable objects) scattering operator which isolates the contribution of the designable object.
A useful operator for this purpose is the $\mathbb{W}$ operator defined by 
\begin{align}
  &\mathbb{I} = \mathbb{W}\left(\mathbb{I}-
  \mathbb{V}\mathbb{G}_{\vac}\right) = 
  \left(\mathbb{I}-\mathbb{V}\mathbb{G}_{\vac}\right)\mathbb{W}. 
  \label{Wformal}
\end{align}
From Eqs.~\eqref{Wformal} and \eqref{eq:clk_Tdefintion}, it is clear that $\mathbb{I} = \mathbb{W}\mathbb{V}\mathbb{T}^{-1}$ and $\mathbb{W} = \mathbb{I} + \mathbb{T}\mathbb{G}_{\vac}$.
The $\mathbb{T}$ operator takes an incident field and gives $\propto$ the generated current, $|\mathbf{J}_{g}\rangle = -\frac{ik_{0}}{Z} \mathbb{T}|\mathbf{E}_{i}\rangle$, while the
$\mathbb{W}$ operator takes an initial current and gives the total current, $|\mathbf{J}_{t}\rangle = \mathbb{T}\mathbb{V}^{-1}|\mathbf{J}_{i}\rangle = \mathbb{W}|\mathbf{J}_{i}\rangle$.
We will make use of the nesting property of the $\mathbb{W}$ operator 
\begin{align}
  \mathbb{W}_{\mathrm{tot}} = \mathbb{W}_{\obj}\mathbb{W}_{\dobj}, 
  \label{wNest}
\end{align}
where $\mathbb{W}_{\mathrm{tot}}$ is the $\mathbb{W}$ operator of the total system (including all the scatterers), $\mathbb{W}_{\obj}$ is the $\mathbb{W}$ operator of the fixed object only, and $\mathbb{W}_{\dobj}$ is the $\mathbb{W}$ operator for the designable body with the fixed object contained in the background, i.e., $\mathbb{G}_{\vac}$ replaced by $\mathbb{G}_{\vac}\mathbb{W}_{\obj} = \mathbb{G}_{\vac}(\mathbb{I} + \mathbb{T}_{\obj}\mathbb{G}_{\vac}) = \mathbb{G}_{\vac} + \mathbb{G}_{\vac}\mathbb{T}_{\obj}\mathbb{G}_{\vac}$ (which is the total Green's function of the fixed object in the background, $\mathbb{G}_{\obj}$) in the defining relation for $\mathbb{W}_{\dobj}$. 
The nesting property holds since $\mathbb{W}_{\dobj}^{-1}\mathbb{W}_{\obj}^{-1} = (\mathbb{I} -\mathbb{V}_{\clk}\mathbb{G}_{\obj})\mathbb{W}_{\obj}^{-1} = (\mathbb{I} - \mathbb{V}_{\clk}\mathbb{G}_{\vac}\mathbb{W}_{\obj})\mathbb{W}_{\obj}^{-1} = (\mathbb{W}_{\obj}^{-1} - \mathbb{V}_{\clk}\mathbb{G}_{\vac}) = (\mathbb{I} - \mathbb{V}_{\obj}\mathbb{G}_{\vac} - \mathbb{V}_{\clk}\mathbb{G}_{\vac}) = \mathbb{W}_{\mathrm{tot}}^{-1}$.
Using $\mathbb{W}_{\mathrm{tot}}\mathbb{V}_{\mathrm{tot}} = \mathbb{T}_{\mathrm{tot}}$ and the nesting relation Eq.~\eqref{wNest} yields
\begin{align}
    \mathbb{T}_{\mathrm{tot}} &= \mathbb{W}_{\mathrm{tot}}\mathbb{V}_{\mathrm{tot}}
    = \mathbb{W}_{\obj}\mathbb{W}_{\dobj}\mathbb{V}_{\mathrm{tot}} \\
    &=
    \mathbb{W}_{\obj}\left(\mathbb{W}_{\dobj}\mathbb{V}_{\obj} + \mathbb{W}_{\dobj}\mathbb{V}_{\clk}\right)
    \\
    &= \mathbb{W}_{\obj}\left(\mathbb{W}_{\dobj}\mathbb{V}_{\obj} + \mathbb{T}_{\dobj}\right)
    \\
    &= \mathbb{W}_{\obj}\left(\mathbb{V}_{\obj} + \mathbb{T}_{\dobj}\left(\mathbb{G}_{\vac}\mathbb{T}_{\obj}+\mathbb{I}\right)\right) \\
    &= \mathbb{T}_{\obj} + \mathbb{G}_{\vac}^{-1}\mathbb{G}_{\obj}\mathbb{T}_{\dobj}\mathbb{G}_{\obj}\mathbb{G}_{\vac}^{-1},
  \label{tNest}
\end{align}
where we made use of $\mathbb{W}_{\dobj} = \mathbb{I} + \mathbb{T}_{\dobj}\mathbb{G}_{\obj} = \mathbb{I} + \mathbb{T}_{\dobj}\mathbb{G}_{\vac}\mathbb{W}_{\obj} = \mathbb{I} + \mathbb{T}_{\dobj}\mathbb{G}_{\vac}\mathbb{T}_{\obj}\mathbb{V}_{\obj}^{-1}$.
Define $\left|\textbf{E}_{\dobj}\right>\equiv \left(\mathbb{G}_{\vac}\mathbb{T}_{\obj} + \mathbb{I}\right)\left|\mathbf{E}_{i}\right> = \frac{iZ}{k_{0}}\mathbb{G}_{\obj}\left|\mathbf{J}_{i}\right>$ and $\left|\textbf{T}_{\dobj}\right>\equiv \mathbb{T}_{\dobj}\left|\textbf{E}_{\dobj}\right>$.
Use $\mathbb{T}_{\mathrm{tot}} = \mathbb{W}_{\obj}(\mathbb{V}_{\obj} + \mathbb{T}_{\dobj}(\mathbb{G}_{\vac}\mathbb{T}_{\obj} + \mathbb{I}))$ to find that the total extinct power by all the scatterers, relative to a vacuum background, is given by
\begin{align}
    P_{\mathrm{ext}} &=
    \frac{1}{2Z}\im{k_{0}\left<\mathbf{E}_{i}\right| \mathbb{T}_{\mathrm{tot}} \left|\mathbf{E}_{i}\right>} \\
    &= \frac{1}{2Z}\im{k_{0}\left<\mathbf{E}_{i}\right|\mathbb{W}_{\obj}\mathbb{V}_{\obj}\left|\mathbf{E}_{i}\right>} \nonumber \\
    &~~+ \frac{1}{2Z}\im{k_{0}\left<\mathbf{E}_{i}\right|\mathbb{W}_{\obj}\mathbb{T}_{\dobj}(\mathbb{G}_{\vac}\mathbb{T}_{\obj} + \mathbb{I})\left|\mathbf{E}_{i}\right>}  \\
    &= \frac{1}{2Z}\im{k_{0}\left<\mathbf{E}_{i}\right|\mathbb{W}_{\obj}\mathbb{V}_{\obj}\left|\mathbf{E}_{i}\right>} \nonumber \\
    &~~+ \frac{1}{2Z}\im{k_{0}\left<\mathbf{E}_{i}\right|\mathbb{W}_{\obj}\left|\textbf{T}_{\dobj}\right>}
    \label{eq:objDressed}
\end{align}
with $\mathbb{T}_{\dobj}$ subject to the fundamental relation 
\begin{align}
    \mathbb{I}_{\des} &= 
    \mathbb{I}_{\des}\left(\chi_{\obj}^{-1}\mathbb{I} - \mathbb{G}_{\obj}\right)\mathbb{I}_{\des}\mathbb{T}_{\dobj},
    \label{tFundaDress}
\end{align}
where $\chi_{\clk}$ is the susceptibility used for the object in the design domain (the cloak).
This constraint is similar to that in previous works~\cite{molesky_global_2020,chao_maximum_2022}, but where the relevant Green's function is $\mathbb{G}_{\obj}$ and not the vacuum Green's function.
Use of $\mathbb{W}_{\obj} = \mathbb{G}_{\vac}^{-1}\mathbb{G}_{\obj}$ leads to the primal objective function presented in the main text.

\section{Topology optimization procedure}
\label{sec:app_top_opt}

To calculate inverse designs, we made use of the NLopt package~\cite{johnson_nlopt_2019} and followed standard topology optimization algorithms~\cite{molesky_inverse_2018,christiansen_inverse_2021} based on the method of moving asymptotes~\cite{svanberg_class_2002} where the electric susceptibility value of each pixel in the specified design region is considered as an independent design parameter.
Each pixel in the design region is allowed to explore a continuous range of susceptibility values varying linearly between the vacuum value $0$ and the prescribed material value $\chi_{\mathrm{clk}}$.
The continuous range $[0, 1] \cdot \chi_{\mathrm{clk}}$ can be interpreted, after normalization by $\chi_{\mathrm{clk}}$, as a filling fraction of the pixel.

We implemented a 2D Maxwell solver which could be used for arbitrary structuring (allow the pixels within any arbitrary 2D design region to vary).
Doing a thousand function evaluations (iterations of optimization over the susceptibility profile for fixed material parameters, design region, bandwidth, etc.) in 2D takes less than 1 hour for a wavelength-scale design region.
The primary challenge in such optimization problems lies in the computational cost of evaluating the objective function and gradient, which is evaluated several hundred if not thousand times during the course of any one optimization.
Let $\mathbb{G}$ denote the total Green's function and let $\mathbb{M}^{-1} = \mathbb{G}$ where $\mathbb{M} = \frac{c^{2}}{\omega_{0}^{2}}\nabla\times\nabla - \epsilon(\mathbf{r})$.
After the discretization of the computational grid, then in the design region $\epsilon_{b} = 1 + \chi_{\clk}\bar{\chi}_{b}$ at pixel $b$. 
The gradient of the objective function $P_{\mathrm{ext}}(\{\bar{\chi}_{b}\};\omega_{0}) =
    \frac{1}{2}\re{\left<\mathbf{E}_{i}|\mathbf{J}_{g}\right>} = \frac{1}{2}\re{\left<\mathbf{E}_{i}|\mathbb{V}\mathbb{G}|\mathbf{J}_{i}\right>}$ with respect to the degree of freedom $\bar{\chi}_{a}$ is given by
\begin{align}
    \frac{\partial P_{\mathrm{ext}}}{\partial \bar{\chi}_{a}}&=
    \frac{1}{2}\re{\left<\mathbf{E}_{i}|\frac{\partial(\mathbb{V}\mathbb{G})}{\partial \bar{\chi}_{a}}|\mathbf{J}_{i}\right>}
    \\
    &=
    \frac{1}{2}\re{\left<\mathbf{E}_{i}|
    (\chi_{\clk}\mathbb{P}_{a}\mathbb{G} + \mathbb{V} \frac{\partial \mathbb{M}^{-1}}{\partial \bar{\chi}_{a}})
    |\mathbf{J}_{i}\right>} \\
    &=
    \frac{1}{2}\re{\left<\mathbf{E}_{i}|
    (\chi_{\clk}\mathbb{P}_{a}\mathbb{G} - \mathbb{V} \mathbb{M}^{-1}\frac{\partial \mathbb{M}}{\partial \bar{\chi}_{a}}\mathbb{M}^{-1})
    |\mathbf{J}_{i}\right>} \\
    &=
    \frac{1}{2}\re{\left<\mathbf{E}_{i}|
    (\chi_{\clk}\mathbb{P}_{a}\mathbb{G} 
    + \chi_{\clk}\mathbb{V} \mathbb{G}\mathbb{P}_{a}\mathbb{G})
    |\mathbf{J}_{i}\right>} \\
    &=
    \frac{1}{2}\re{\frac{k_{0}\chi_{\clk}}{iZ}\left<\mathbf{E}_{i}|
    (\mathbb{P}_{a} + \mathbb{V} \mathbb{G}\mathbb{P}_{a})
    |\mathbf{E}_{t}\right>} \\
    &\equiv
    \frac{1}{2}\re{\frac{k_{0}\chi_{\clk}}{iZ}\langle\mathbf{E}_{i} + \tilde{\mathbf{E}}| 
    \mathbb{P}_{a} |\mathbf{E}_{t}\rangle}
    \label{eq:Pextcloakinggradient}
\end{align}
where $\bar{\chi}_{a}$ is a topology optimization degree of freedom, normalized to take values in $[0, 1]$, so that the susceptibility at the $a$-th pixel in the design region is given by $\chi_{\mathrm{clk}}\bar{\chi}_{a}$, $\mathbb{P}_{a}$ is projection onto the pixel indexed by $a$, $|\mathbf{E}_{t}\rangle$ is the total electric field, and $|\tilde{\mathbf{E}}\rangle \equiv \mathbb{G}^{\dagger}\mathbb{V}^{\dagger}|\mathbf{E}_{i}\rangle$ which for reciprocal systems becomes $|\tilde{\mathbf{E}}\rangle \equiv (\mathbb{G}\mathbb{V}|\mathbf{E}_{i}^{*}\rangle)^{*}$.
Hence, for a given incident field $|\mathbf{E}_{i}\rangle$ one only needs to perform two additional Maxwell solves to calculate $|\mathbf{E}_{t}\rangle$ and $|\tilde{\mathbf{E}}\rangle$ at each iteration of the optimization to obtain the required gradient Eq.~\eqref{eq:Pextcloakinggradient} and the objective function
$P_{\mathrm{ext}}(\{\bar{\chi}_{b}\};\omega_{0})
=
\frac{1}{2}\re{\left<\mathbf{E}_{i}|\mathbf{J}_{g}\right>}
=
\frac{1}{2Z}\im{k_{0}\left<\mathbf{E}_{i}|\mathbb{V}|\mathbf{E}_{t}\right>}$.
Details on the computational complexities encountered in 3D and potential solution methods can be found in Appendix C of Ref.~\cite{molesky_global_2020} and the therein cited references.

%REFERENCES
\bibliography{refs}

\begin{thebibliography}{45}
\providecommand{\natexlab}[1]{#1}
\providecommand{\url}[1]{\texttt{#1}}
\expandafter\ifx\csname urlstyle\endcsname\relax
  \providecommand{\doi}[1]{doi: #1}\else
  \providecommand{\doi}{doi: \begingroup \urlstyle{rm}\Url}\fi

\bibitem[Nicolai and Carichner(2010)]{nicolai_fundamentals_2010}
Leland~M. Nicolai and Grant~E. Carichner.
\newblock \emph{Fundamentals of Aircraft and Airship Design: Volume I--Aircraft Design}.
\newblock American Institute of Aeronautics and Astronautics, Inc., 2010.

\bibitem[Li and Pendry(2008)]{li_hiding_carpet_2008}
Jensen Li and John~B. Pendry.
\newblock Hiding under the carpet: a new strategy for cloaking.
\newblock \emph{Physical Review Letters}, 101\penalty0 (20):\penalty0 203901, 2008.

\bibitem[Ergin et~al.(2010)Ergin, Stenger, Brenner, Pendry, and Wegener]{ergin_3D_cloak_2010}
Tolga Ergin, Nicolas Stenger, Patrice Brenner, John~B. Pendry, and Martin Wegener.
\newblock Three-dimensional invisibility cloak at optical wavelengths.
\newblock \emph{Science}, 328\penalty0 (5976):\penalty0 337--339, 2010.

\bibitem[Al{\`u} and Engheta(2009)]{alu_cloaking_sensor_2009}
Andrea Al{\`u} and Nader Engheta.
\newblock Cloaking a sensor.
\newblock \emph{Physical Review Letters}, 102\penalty0 (23):\penalty0 233901, 2009.

\bibitem[Pendry(2006)]{pendry_controlling_2006}
J.~B. Pendry.
\newblock Controlling electromagnetic fields.
\newblock \emph{Science}, 312\penalty0 (5781):\penalty0 1780--1782, June 2006.
\newblock ISSN 0036-8075, 1095-9203.
\newblock \doi{10.1126/science.1125907}.

\bibitem[Al{\`u} and Engheta(2005)]{alu_plasmonic_2005}
Andrea Al{\`u} and Nader Engheta.
\newblock Achieving transparency with plasmonic and metamaterial coatings.
\newblock \emph{Physical Review E}, 72\penalty0 (1):\penalty0 016623, 2005.

\bibitem[Al{\`u}(2009)]{alu_mantle_2009}
Andrea Al{\`u}.
\newblock Mantle cloak: Invisibility induced by a surface.
\newblock \emph{Physical Review B}, 80\penalty0 (24):\penalty0 245115, 2009.

\bibitem[Tretyakov et~al.(2009)Tretyakov, Alitalo, Luukkonen, and Simovski]{tretyakov_broadband_2009}
Sergei Tretyakov, Pekka Alitalo, Olli Luukkonen, and Constantin Simovski.
\newblock Broadband electromagnetic cloaking of long cylindrical objects.
\newblock \emph{Physical Review Letters}, 103\penalty0 (10):\penalty0 103905, 2009.

\bibitem[Alitalo et~al.(2008)Alitalo, Luukkonen, Jylha, Venermo, and Tretyakov]{alitalo_transmission_2008}
Pekka Alitalo, Olli Luukkonen, Liisi Jylha, Jukka Venermo, and Sergei~A. Tretyakov.
\newblock Transmission-line networks cloaking objects from electromagnetic fields.
\newblock \emph{IEEE Transactions on Antennas and Propagation}, 56\penalty0 (2):\penalty0 416--424, 2008.

\bibitem[Milton and Nicorovici(2006)]{milton_cloaking_2006}
Graeme~W. Milton and Nicolae-Alexandru~P. Nicorovici.
\newblock On the cloaking effects associated with anomalous localized resonance.
\newblock \emph{Proceedings of the Royal Society A: Mathematical, Physical and Engineering Sciences}, 462\penalty0 (2074):\penalty0 3027--3059, 2006.

\bibitem[Miller(2006)]{miller_perfect_2006}
David A.~B. Miller.
\newblock On perfect cloaking.
\newblock \emph{Optics Express}, 14\penalty0 (25):\penalty0 12457--12466, 2006.

\bibitem[Hashemi et~al.(2012)Hashemi, Qiu, McCauley, Joannopoulos, and Johnson]{hashemi_diameter-bandwidth_2012}
Hila Hashemi, Cheng-Wei Qiu, Alexander~P. McCauley, J.~D. Joannopoulos, and Steven~G. Johnson.
\newblock Diameter-bandwidth product limitation of isolated-object cloaking.
\newblock \emph{Physical Review A}, 86\penalty0 (1):\penalty0 013804, July 2012.
\newblock ISSN 1050-2947, 1094-1622.
\newblock \doi{10.1103/PhysRevA.86.013804}.

\bibitem[Chao et~al.(2022{\natexlab{a}})Chao, Strekha, Kuate~Defo, Molesky, and Rodriguez]{chao_physical_2022}
Pengning Chao, Benjamin Strekha, Rodrick Kuate~Defo, Sean Molesky, and Alejandro~W. Rodriguez.
\newblock Physical limits in electromagnetism.
\newblock \emph{Nature Reviews Physics}, 4\penalty0 (8):\penalty0 543--559, August 2022{\natexlab{a}}.
\newblock ISSN 2522-5820.
\newblock \doi{10.1038/s42254-022-00468-w}.

\bibitem[Molesky et~al.(2020{\natexlab{a}})Molesky, Chao, Jin, and Rodriguez]{molesky_global_2020}
Sean Molesky, Pengning Chao, Weiliang Jin, and Alejandro~W. Rodriguez.
\newblock Global {$\mathbb{T}$} operator bounds on electromagnetic scattering: Upper bounds on far-field cross sections.
\newblock \emph{Physical Review Research}, 2\penalty0 (3):\penalty0 033172, 2020{\natexlab{a}}.

\bibitem[Jelinek et~al.(2021)Jelinek, Gustafsson, Capek, and Schab]{jelinek_fundamental_2021}
Lukas Jelinek, Mats Gustafsson, Miloslav Capek, and Kurt Schab.
\newblock Fundamental bounds on the performance of monochromatic passive cloaks.
\newblock \emph{Optics Express}, 29\penalty0 (15):\penalty0 24068, July 2021.
\newblock ISSN 1094-4087.
\newblock \doi{10.1364/OE.428536}.

\bibitem[Cai et~al.(2007)Cai, Chettiar, Kildishev, and Shalaev]{cai_optical_cloaking_2007}
Wenshan Cai, Uday~K. Chettiar, Alexander~V. Kildishev, and Vladimir~M. Shalaev.
\newblock Optical cloaking with metamaterials.
\newblock \emph{Nature Photonics}, 1\penalty0 (4):\penalty0 224--227, 2007.

\bibitem[Schurig et~al.(2006)Schurig, Pendry, and Smith]{schurig_calculation_2006}
David Schurig, John~B. Pendry, and David~R. Smith.
\newblock Calculation of material properties and ray tracing in transformation media.
\newblock \emph{Optics Express}, 14\penalty0 (21):\penalty0 9794--9804, 2006.

\bibitem[Monticone and Al{\`u}(2016)]{monticone_invisibility_2016}
Francesco Monticone and Andrea Al{\`u}.
\newblock Invisibility exposed: Physical bounds on passive cloaking.
\newblock \emph{Optica}, 3\penalty0 (7):\penalty0 718--724, 2016.

\bibitem[Cassier and Milton(2017)]{cassier_bounds_2017}
Maxence Cassier and Graeme~W. Milton.
\newblock Bounds on {{Herglotz}} functions and fundamental limits of broadband passive quasistatic cloaking.
\newblock \emph{Journal of Mathematical Physics}, 58\penalty0 (7):\penalty0 071504, 2017.

\bibitem[Jackson(1999)]{jackson_classical_1999}
John~David Jackson.
\newblock \emph{Classical Electrodynamics}.
\newblock {Wiley}, {New York}, 3rd edition, 1999.
\newblock ISBN 978-0-471-30932-1.

\bibitem[Kuang and Miller(2020)]{kuang_computational_2020}
Zeyu Kuang and Owen~D. Miller.
\newblock Computational bounds to light--matter interactions via local conservation laws.
\newblock \emph{Physical Review Letters}, 125\penalty0 (26):\penalty0 263607, December 2020.
\newblock \doi{10.1103/PhysRevLett.125.263607}.

\bibitem[Molesky et~al.(2020{\natexlab{b}})Molesky, Venkataram, Jin, and Rodriguez]{molesky_fundamental_2020}
Sean Molesky, Prashanth~S. Venkataram, Weiliang Jin, and Alejandro~W. Rodriguez.
\newblock Fundamental limits to radiative heat transfer: Theory.
\newblock \emph{Physical Review B}, 101\penalty0 (3):\penalty0 035408, 2020{\natexlab{b}}.

\bibitem[Venkataram et~al.(2020{\natexlab{a}})Venkataram, Molesky, Jin, and Rodriguez]{venkataram_fundamental_2020-1}
Prashanth~S. Venkataram, Sean Molesky, Weiliang Jin, and Alejandro~W. Rodriguez.
\newblock Fundamental limits to radiative heat transfer: The limited role of nanostructuring in the near-field.
\newblock \emph{Physical Review Letters}, 124\penalty0 (1):\penalty0 013904, 2020{\natexlab{a}}.

\bibitem[Novotny and Hecht(2012)]{novotny_principles_2012}
Lukas Novotny and Bert Hecht.
\newblock \emph{Principles of Nano-Optics}.
\newblock {Cambridge University Press}, {Cambridge}, second edition, 2012.
\newblock ISBN 978-1-139-55425-1.

\bibitem[Boyd and Vandenberghe(2004)]{boyd_convex_2004}
Stephen~P. Boyd and Lieven Vandenberghe.
\newblock \emph{Convex Optimization}.
\newblock {Cambridge University Press}, {Cambridge, UK; New York}, 2004.
\newblock ISBN 978-0-521-83378-3.

\bibitem[Molesky et~al.(2018)Molesky, Lin, Piggott, Jin, Vu{\v c}kovi{\'c}, and Rodriguez]{molesky_inverse_2018}
Sean Molesky, Zin Lin, Alexander~Y. Piggott, Weiliang Jin, Jelena Vu{\v c}kovi{\'c}, and Alejandro~W. Rodriguez.
\newblock Inverse design in nanophotonics.
\newblock \emph{Nature Photonics}, 12\penalty0 (11):\penalty0 659--670, 2018.

\bibitem[Molesky et~al.(2020{\natexlab{c}})Molesky, Chao, and Rodriguez]{molesky_hierarchical_2020}
Sean Molesky, Pengning Chao, and Alejandro~W. Rodriguez.
\newblock Hierarchical mean-field {$\mathbb{T}$} operator bounds on electromagnetic scattering: {{Upper}} bounds on near-field radiative {{Purcell}} enhancement.
\newblock \emph{Physical Review Research}, 2\penalty0 (4):\penalty0 043398, December 2020{\natexlab{c}}.
\newblock \doi{10.1103/PhysRevResearch.2.043398}.

\bibitem[Chao et~al.(2022{\natexlab{b}})Chao, Kuate~Defo, Molesky, and Rodriguez]{chao_maximum_2022}
Pengning Chao, Rodrick Kuate~Defo, Sean Molesky, and Alejandro Rodriguez.
\newblock Maximum electromagnetic local density of states via material structuring.
\newblock \emph{Nanophotonics}, 12\penalty0 (3):\penalty0 549--557, 2022{\natexlab{b}}.

\bibitem[Angeris et~al.(2019)Angeris, Vu{\v c}kovi{\'c}, and Boyd]{angeris_computational_2019}
Guillermo Angeris, Jelena Vu{\v c}kovi{\'c}, and Stephen~P. Boyd.
\newblock Computational bounds for photonic design.
\newblock \emph{ACS Photonics}, 6\penalty0 (5):\penalty0 1232, 2019.
\newblock \doi{10.1021/acsphotonics.9b00154}.

\bibitem[Shim et~al.(2019)Shim, Fan, Johnson, and Miller]{shim_fundamental_2019}
Hyungki Shim, Lingling Fan, Steven~G. Johnson, and Owen~D. Miller.
\newblock Fundamental {{Limits}} to {{Near-Field Optical Response}} over {{Any Bandwidth}}.
\newblock \emph{Physical Review X}, 9\penalty0 (1):\penalty0 011043, March 2019.
\newblock ISSN 2160-3308.
\newblock \doi{10.1103/PhysRevX.9.011043}.

\bibitem[Amaolo et~al.(2024{\natexlab{a}})Amaolo, Chao, Maldonado, Molesky, and Rodriguez]{amaolo_can_heterostructures_2023}
Alessio Amaolo, Pengning Chao, Thomas~J. Maldonado, Sean Molesky, and Alejandro~W. Rodriguez.
\newblock Can photonic heterostructures provably outperform single-material geometries?
\newblock \emph{Nanophotonics}, 13\penalty0 (3):\penalty0 283--288, 2024{\natexlab{a}}.

\bibitem[Mohajan et~al.(2023)Mohajan, Chao, Jin, Molesky, and Rodriguez]{mohajan_fundamental_2023}
Jewel Mohajan, Pengning Chao, Weiliang Jin, Sean Molesky, and Alejandro~W. Rodriguez.
\newblock Fundamental limits on radiative $\chi$ (2) second harmonic generation.
\newblock \emph{Optics Express}, 31\penalty0 (26):\penalty0 44212--44223, 2023.

\bibitem[Strekha et~al.(2024{\natexlab{a}})Strekha, Chao, Defo, Molesky, and Rodriguez]{strekha_suppressing_2024}
Benjamin Strekha, Pengning Chao, Rodrick~Kuate Defo, Sean Molesky, and Alejandro~W. Rodriguez.
\newblock Suppressing electromagnetic local density of states via slow light in lossy quasi-one-dimensional gratings.
\newblock \emph{Physical Review A}, 109\penalty0 (4):\penalty0 L041501, 2024{\natexlab{a}}.

\bibitem[Amaolo et~al.(2024{\natexlab{b}})Amaolo, Chao, Maldonado, Molesky, and Rodriguez]{amaolo_raman_2024}
Alessio Amaolo, Pengning Chao, Thomas~J. Maldonado, Sean Molesky, and Alejandro~W. Rodriguez.
\newblock Physical limits on {R}aman scattering: The critical role of pump and signal co-design.
\newblock \emph{arXiv preprint arXiv:2403.03332}, 2024{\natexlab{b}}.

\bibitem[Venkataram et~al.(2020{\natexlab{b}})Venkataram, Molesky, Chao, and Rodriguez]{venkataram_fundamental_2020-2}
Prashanth~S. Venkataram, Sean Molesky, Pengning Chao, and Alejandro~W. Rodriguez.
\newblock Fundamental limits to attractive and repulsive {{Casimir-Polder}} forces.
\newblock \emph{Physical Review A}, 101\penalty0 (5):\penalty0 052115, 2020{\natexlab{b}}.

\bibitem[Strekha et~al.(2022)Strekha, Molesky, Chao, Kr{\"u}ger, and Rodriguez]{strekha_tracenoneq_2022}
Benjamin Strekha, Sean Molesky, Pengning Chao, Matthias Kr{\"u}ger, and Alejandro~W. Rodriguez.
\newblock Trace expressions and associated limits for nonequilibrium {C}asimir torque.
\newblock \emph{Physical Review A}, 106\penalty0 (4):\penalty0 042222, 2022.

\bibitem[Strekha et~al.(2024{\natexlab{b}})Strekha, Kr{\"u}ger, and Rodriguez]{strekha_traceeq_2024}
Benjamin Strekha, Matthias Kr{\"u}ger, and Alejandro~W. Rodriguez.
\newblock Trace expressions and associated limits for equilibrium {C}asimir torque.
\newblock \emph{Physical Review A}, 109\penalty0 (1):\penalty0 012813, 2024{\natexlab{b}}.

\bibitem[Fleury et~al.(2015)Fleury, Monticone, and Al{\`u}]{fleury_invisibility_2015}
Romain Fleury, Francesco Monticone, and Andrea Al{\`u}.
\newblock Invisibility and cloaking: Origins, present, and future perspectives.
\newblock \emph{Physical Review Applied}, 4\penalty0 (3):\penalty0 037001, 2015.
\newblock \doi{10.1103/PhysRevApplied.4.037001}.

\bibitem[Monticone and Al{\`u}(2013)]{monticone_cloaked_2013}
Francesco Monticone and Andrea Al{\`u}.
\newblock Do cloaked objects really scatter less?
\newblock \emph{Physical Review X}, 3\penalty0 (4):\penalty0 041005, 2013.

\bibitem[Hashemi et~al.(2011)Hashemi, Oskooi, Joannopoulos, and Johnson]{hashemi_general_2011}
Hila Hashemi, A.~Oskooi, J.~D. Joannopoulos, and Steven~G. Johnson.
\newblock General scaling limitations of ground-plane and isolated-object cloaks.
\newblock \emph{Physical Review A}, 84\penalty0 (2):\penalty0 023815, August 2011.
\newblock ISSN 1050-2947, 1094-1622.
\newblock \doi{10.1103/PhysRevA.84.023815}.

\bibitem[Molesky et~al.(2021)Molesky, Chao, Mohajan, Reinhart, Chi, and Rodriguez]{molesky_mathbbt-operator_2021}
Sean Molesky, Pengning Chao, Jewel Mohajan, Wesley Reinhart, Heng Chi, and Alejandro~W. Rodriguez.
\newblock {$\mathbb{T}$}-{{Operator Limits}} on {{Optical Communication}}: {{Metaoptics}}, {{Computation}}, and {{Input-Output Transformations}}.
\newblock \emph{Physical Review Research}, 2021.

\bibitem[Shim et~al.(2024)Shim, Kuang, Lin, and Miller]{shim_multi-functional_2024}
Hyungki Shim, Zeyu Kuang, Zin Lin, and Owen~D. Miller.
\newblock Fundamental limits to multi-functional and tunable nanophotonic response.
\newblock \emph{Nanophotonics}, 13\penalty0 (12):\penalty0 2107--2116, 2024.

\bibitem[Johnson et~al.(2019)]{johnson_nlopt_2019}
S.~G. Johnson et~al.
\newblock \emph{The {{NLopt}} Nonlinear Optimization Package (Version 2.6.2)}.
\newblock August 2019.

\bibitem[Christiansen and Sigmund(2021)]{christiansen_inverse_2021}
Rasmus~E. Christiansen and Ole Sigmund.
\newblock Inverse design in photonics by topology optimization: Tutorial.
\newblock \emph{JOSA B}, 38\penalty0 (2):\penalty0 496--509, 2021.

\bibitem[Svanberg(2002)]{svanberg_class_2002}
Krister Svanberg.
\newblock A class of globally convergent optimization methods based on conservative convex separable approximations.
\newblock \emph{SIAM Journal on Optimization}, 12\penalty0 (2):\penalty0 555, 2002.

\end{thebibliography}


\begin{thebibliography}{9}
\providecommand{\natexlab}[1]{#1}
\providecommand{\url}[1]{\texttt{#1}}
\expandafter\ifx\csname urlstyle\endcsname\relax
  \providecommand{\doi}[1]{doi: #1}\else
  \providecommand{\doi}{doi: \begingroup \urlstyle{rm}\Url}\fi

\bibitem[Molesky et~al.(2020{\natexlab{a}})Molesky, Chao, Jin, and Rodriguez]{molesky_global_2020}
Sean Molesky, Pengning Chao, Weiliang Jin, and Alejandro~W. Rodriguez.
\newblock Global {$\mathbb{T}$} operator bounds on electromagnetic scattering: Upper bounds on far-field cross sections.
\newblock \emph{Physical Review Research}, 2\penalty0 (3):\penalty0 033172, 2020{\natexlab{a}}.

\bibitem[Molesky et~al.(2020{\natexlab{b}})Molesky, Venkataram, Jin, and Rodriguez]{molesky_fundamental_2020}
Sean Molesky, Prashanth~S. Venkataram, Weiliang Jin, and Alejandro~W. Rodriguez.
\newblock Fundamental limits to radiative heat transfer: Theory.
\newblock \emph{Physical Review B}, 101\penalty0 (3):\penalty0 035408, 2020{\natexlab{b}}.

\bibitem[Venkataram et~al.(2020)Venkataram, Molesky, Jin, and Rodriguez]{venkataram_fundamental_2020-1}
Prashanth~S. Venkataram, Sean Molesky, Weiliang Jin, and Alejandro~W. Rodriguez.
\newblock Fundamental limits to radiative heat transfer: The limited role of nanostructuring in the near-field.
\newblock \emph{Physical Review Letters}, 124\penalty0 (1):\penalty0 013904, 2020.

\bibitem[Tsang et~al.(2004)Tsang, Kong, and Ding]{tsang_scattering_2004}
Leung Tsang, Jin~Au Kong, and Kung-Hau Ding.
\newblock \emph{Scattering of Electromagnetic Waves: Theories and Applications}, volume~27.
\newblock {John Wiley \& Sons}, 2004.

\bibitem[Jackson(1999)]{jackson_classical_1999}
John~David Jackson.
\newblock \emph{Classical Electrodynamics}.
\newblock {Wiley}, {New York}, 3rd edition, 1999.
\newblock ISBN 978-0-471-30932-1.

\bibitem[Novotny and Hecht(2012)]{novotny_principles_2012}
Lukas Novotny and Bert Hecht.
\newblock \emph{Principles of Nano-Optics}.
\newblock {Cambridge University Press}, {Cambridge}, second edition, 2012.
\newblock ISBN 978-1-139-55425-1.

\bibitem[Chao et~al.(2022)Chao, Kuate~Defo, Molesky, and Rodriguez]{chao_maximum_2022}
Pengning Chao, Rodrick Kuate~Defo, Sean Molesky, and Alejandro Rodriguez.
\newblock Maximum electromagnetic local density of states via material structuring.
\newblock \emph{Nanophotonics}, 12\penalty0 (3):\penalty0 549--557, 2022.

\bibitem[Strekha et~al.(2024)Strekha, Chao, Defo, Molesky, and Rodriguez]{strekha_suppressing_2024}
Benjamin Strekha, Pengning Chao, Rodrick~Kuate Defo, Sean Molesky, and Alejandro~W. Rodriguez.
\newblock Suppressing electromagnetic local density of states via slow light in lossy quasi-one-dimensional gratings.
\newblock \emph{Physical Review A}, 109\penalty0 (4):\penalty0 L041501, 2024.

\bibitem[Boyd and Vandenberghe(2004)]{boyd_convex_2004}
Stephen~P. Boyd and Lieven Vandenberghe.
\newblock \emph{Convex Optimization}.
\newblock {Cambridge University Press}, {Cambridge, UK; New York}, 2004.
\newblock ISBN 978-0-521-83378-3.

\end{thebibliography}
\end{document}

\title{Limitations on bandwidth-integrated passive cloaking: supplemental material}

\author{Benjamin Strekha}
\affiliation{Department of Electrical and Computer Engineering, Princeton University, Princeton, New Jersey 08544, USA}

\author{Alessio Amaolo}
\affiliation{Department of Chemistry, Princeton University, Princeton, New Jersey 08544, USA}

\author{Jewel Mohajan}
\affiliation{Department of Electrical and Computer Engineering, Princeton University, Princeton, New Jersey 08544, USA}

\author{Pengning Chao}
\affiliation{Department of Mathematics, Massachusetts Institute of Technology, Cambridge, Massachusetts 02139, USA}

\author{Sean Molesky}
\affiliation{Department of Engineering Physics, Polytechnique Montréal, Montréal, Québec H3T 1J4, Canada}

\author{Alejandro W. Rodriguez}
\affiliation{Department of Electrical and Computer Engineering, Princeton University, Princeton, New Jersey 08544, USA}
% \email{arod@princeton.edu}

\begin{abstract}
    This set of notes explores the introduction of multiregional device descriptions using the scattering $\mathbb{T}$ operator formalism and formulates objectives for use in a Lagrange dual framework for computing bounds. 
    Applications to extinguished power are considered. 
\end{abstract}

\maketitle

\section{Introduction}

\subsection{Notation} 
Throughout the text, we use the notation found in previous descriptions of $\mathbb{T}$ operator bounds~\cite{molesky_global_2020,molesky_fundamental_2020,venkataram_fundamental_2020-1}. 
Following conventional scattering theory~\cite{tsang_scattering_2004}, an \emph{initial (incident, given, or bare) field} is denoted with a subscript $i$ (either $\left|\mathbf{E}_{i}\right>$ or $\left|\mathbf{J}_{i}\right>$) and a \emph{total (or dressed) field} is denoted with a subscript $t$. 
For a pair of initial and total quantities referring to the same underlying field, the \emph{scattered field}, subscript $s$, is defined as the difference $\left|\textbf{F}_{s}\right> = \left|\textbf{F}_{t}\right> - \left|\textbf{F}_{i}\right>.$ 
The total polarization field of an initial flux problem (or total electromagnetic field of an initial source problem) is referred to as a \emph{generated field} and denoted with a subscript $g$.
Under these definitions, scattering theory is summarized as two sets of formal relations, depending on the initial conditions.
\\ \\
\emph{Initial flux} 
\begin{align}
  \left|\mathbf{J}_{g}\right> &= -\frac{ik_{0}}{Z}\mathbb{V}\left|
  \mathbf{E}_{t}\right> & \left|\mathbf{E}_{t}\right> &=
  \mathbb{V}^{-1}\mathbb{T}\left|\mathbf{E}_{i}\right>\nonumber
  \\ \left|\mathbf{E}_{t}\right> &= \left|\mathbf{E}_{i}\right> +
  \frac{iZ}{k_{0}}\mathbb{G}_{0}\left|\mathbf{J}_{g}\right> &
  \left|\mathbf{E}_{s}\right> &=
  \frac{iZ}{k_{0}}\mathbb{G}_{0}\left|\mathbf{J}_{g}\right>
   \label{eq:supp_clk_initFlux}
\end{align}
\emph{Initial source}
\begin{align}
  \left|\mathbf{E}_{g}\right> &=
  \frac{iZ}{k_{0}}\mathbb{G}_{0}\left|\mathbf{J}_{t}\right> &
  \left|\mathbf{J}_{t}\right> &=
  \mathbb{T}\mathbb{V}^{-1}\left|\mathbf{J}_{i}\right>\nonumber
  \\ \left|\mathbf{J}_{t}\right> &= \left|\mathbf{J}_{i}\right> -
  \frac{ik_{0}}{Z}\mathbb{V}\left|\mathbf{E}_{g}\right> &
  \left|\mathbf{J}_{s}\right> &= -\frac{ik_{0}}{Z} \mathbb{V} \left|
  \mathbf{E}_{g}\right>
   \label{eq:supp_clk_initSource}
 \end{align}
In these equations, $\mathbb{G}_{0}$ stands for the \emph{background} or \emph{environmental} Green's function, including an additional factor of $k^{2}_{0} = \left(2\pi/\lambda_{0}\right)^{2}$ compared to the definition given by most authors~\cite{jackson_classical_1999,novotny_principles_2012}, which may or may not be vacuum. 
The $\mathbb{V}$ operator refers to the scattering potential (susceptibility) relative to this background (whatever material was not included in the definition of $\mathbb{G}_{0}$), and $\left|\mathbf{E}_{i}\right>$ and $\left|\mathbf{J}_{i}\right>$ are similarly defined as initial electric fields and electric currents in the background. 
The remaining quantities in Eqs.~\eqref{eq:supp_clk_initFlux} and \eqref{eq:supp_clk_initSource} are the impedance of free space $Z = \sqrt{\mu_{0}/\epsilon_{0}}$ and the $\mathbb{T}$ operator, defined by the relation
\begin{equation}
	\mathbb{I} = \mathbb{T}\left(\mathbb{V}^{-1} - \mathbb{G}_{0}\right) = \left(\mathbb{V}^{-1} - \mathbb{G}_{0}\right)\mathbb{T}.
	\label{eq:supp_clk_Tdefinition}
\end{equation} 
Another useful operator is the $\mathbb{W}$ operator, defined by 
\begin{align}
  &\mathbb{I} = \mathbb{W}\left(\mathbb{I}-
  \mathbb{V}\mathbb{G}_{0}\right) = 
  \left(\mathbb{I}-\mathbb{V}\mathbb{G}_{0}\right)\mathbb{W}. 
  \label{Wformal}
\end{align}
From Eqs.~\eqref{Wformal} and \eqref{eq:supp_clk_Tdefinition}, it is clear that $\mathbb{I} = \mathbb{W}\mathbb{V}\mathbb{T}^{-1}$ and $\mathbb{W} = \mathbb{I} + \mathbb{T}\mathbb{G}_{0}$.
The $\mathbb{T}$ operator takes an incident field and gives $\propto$ the generated current (namely, $|\mathbf{J}_{g}\rangle = -\frac{ik_{0}}{Z} \mathbb{T}|\mathbf{E}_{i}\rangle$), while the
$\mathbb{W}$ operator takes an initial current and gives the total current (namely, $|\mathbf{J}_{t}\rangle = \mathbb{T}\mathbb{V}^{-1}|\mathbf{J}_{i}\rangle = \mathbb{W}|\mathbf{J}_{i}\rangle$).
Lastly, in the notation below we will need to take the real and imaginary parts of complex-valued equations, leading to equations that can be written in terms of $\asym{\Theta} \equiv \frac{\Theta - \Theta^{\dagger}}{2i}$ and $\sym{\Theta} \equiv \frac{\Theta + \Theta^{\dagger}}{2}$, where $\dagger$ denotes conjugate transpose.
For reciprocal systems with $\Theta = \Theta^{T},$ this notation simplifies to $\asym{\Theta} = \im{\Theta} = \frac{\Theta - \Theta^{*}}{2i}$ and $\sym{\Theta} = \re{\Theta} = \frac{\Theta + \Theta^{*}}{2}$.

\subsection{Expression for Lorentzian bandwidth-integrated extinct power}
Applying the $\mathbb{T}$ operator relations in an initial flux setting, 
\begin{equation}
  P_{\mathrm{flx}}^{\text{ext}} = 
  \frac{1}{2}\re{\left<\mathbf{E}_{i}|\mathbf{J}_{g}\right>}
  = 
  \frac{1}{2Z}\im{k_{0}\left<\mathbf{E}_{i}|\mathbb{T}|\mathbf{E}_{i}\right>}
  =
  \frac{1}{2Z}\im{k_{0}\left<\mathbf{E}_{i}|\mathbb{V}|\mathbf{E}_{t}\right>}
  \label{eq:supp_clk_eIncExt}
\end{equation}
where $k_{0}$ is left inside Im because it can be a complex number in bandwidth-averaged problems.

\section{Nested Formulation}
To shorten the notation in the derivation of the optimization problems, let the $f$ subscript denote the fixed object/region and $d$ denote the designable object/region.
One approach in the presence of a fixed object is to shift the known scattering properties of the fixed object into the background. We wish to write the total scattering operators in a way that isolates the contributions from the fixed and designable objects.
To simplify this procedure, we will make use of the nesting property of the $\mathbb{W}$ operator 
\begin{align}
  \mathbb{W}_{\mathrm{tot}} = \mathbb{W}_{f}\mathbb{W}_{\bowtie}, 
  \label{eq:supp_clk_wNest}
\end{align}
where $\mathbb{W}_{\mathrm{tot}}$ is the $\mathbb{W}$ operator of the total system (including all the scatterers), $\mathbb{W}_{f}$ is the $\mathbb{W}$ operator of the fixed object only, and $\mathbb{W}_{\bowtie}$ is the $\mathbb{W}$ operator for 
the designable object (the cloak) dressed (hence the bowtie) by the fixed object contained in the background, i.e., $\mathbb{G}_{0}$ replaced by $\mathbb{G}_{0}\mathbb{W}_{f} = \mathbb{G}_{0}(\mathbb{I} + \mathbb{T}_{f}\mathbb{G}_{0}) = \mathbb{G}_{0} + \mathbb{G}_{0}\mathbb{T}_{f}\mathbb{G}_{0}$ (which is the total Green's function of the fixed object in the background, $\mathbb{G}_{f}$) in the defining relation for $\mathbb{W}_{\bowtie}$. 
The nesting property holds since $\mathbb{W}_{\bowtie}^{-1}\mathbb{W}_{f}^{-1} = (\mathbb{I} -\mathbb{V}_{d}\mathbb{G}_{f})\mathbb{W}_{f}^{-1} = (\mathbb{I} - \mathbb{V}_{d}\mathbb{G}_{0}\mathbb{W}_{f})\mathbb{W}_{f}^{-1} = (\mathbb{W}_{f}^{-1} - \mathbb{V}_{d}\mathbb{G}_{0}) = (\mathbb{I} - \mathbb{V}_{f}\mathbb{G}_{0} - \mathbb{V}_{d}\mathbb{G}_{0}) = \mathbb{W}_{\mathrm{tot}}^{-1}$.
Using $\mathbb{W}_{\mathrm{tot}}\mathbb{V}_{\mathrm{tot}} = \mathbb{T}_{\mathrm{tot}}$ and the nesting relation Eq.~\eqref{eq:supp_clk_wNest} yields
\begin{align}
  \mathbb{T}_{\mathrm{tot}} = \mathbb{W}_{\mathrm{tot}}\mathbb{V}_{\mathrm{tot}}
  = \mathbb{W}_{f}\mathbb{W}_{\bowtie}\mathbb{V}_{\mathrm{tot}}
  =
  \mathbb{W}_{f}\left(\mathbb{W}_{\bowtie}\mathbb{V}_{f} + \mathbb{W}_{\bowtie}\mathbb{V}_{d}\right)
  = \mathbb{W}_{f}\left(\mathbb{W}_{\bowtie}\mathbb{V}_{f} + \mathbb{T}_{\bowtie}\right)
  &= \mathbb{W}_{f}\left(\mathbb{V}_{f} + \mathbb{T}_{\bowtie}\left(\mathbb{G}_{0}\mathbb{T}_{f}+\mathbb{I}\right)\right) \\
  &= \mathbb{T}_{f} + \mathbb{G}_{0}^{-1}\mathbb{G}_{f}\mathbb{T}_{\bowtie}\mathbb{G}_{f}\mathbb{G}_{0}^{-1},
  \label{tNest}
\end{align}
where we made use of $\mathbb{W}_{\bowtie} = \mathbb{I} + \mathbb{T}_{\bowtie}\mathbb{G}_{f} = \mathbb{I} + \mathbb{T}_{\bowtie}\mathbb{G}_{0}\mathbb{W}_{f} = \mathbb{I} + \mathbb{T}_{\bowtie}\mathbb{G}_{0}\mathbb{T}_{f}\mathbb{V}_{f}^{-1}$.
Define $\left|\textbf{E}_{\bowtie}\right>\equiv \left(\mathbb{G}_{0}\mathbb{T}_{f} + \mathbb{I}\right)\left|\mathbf{E}_{i}\right> = \frac{iZ}{k_{0}}\mathbb{G}_{f}\left|\mathbf{J}_{i}\right>$, the total field in the presence of only the fixed object, and $\left|\textbf{T}_{\bowtie}\right>\equiv \mathbb{T}_{\bowtie}\left|\textbf{E}_{\bowtie}\right>$ so that $-\frac{ik_{0}}{Z}|\mathbf{T}_{\bowtie}\rangle$ is the induced current in the designable object.
Use $\mathbb{T}_{\mathrm{tot}} = \mathbb{W}_{f}(\mathbb{V}_{f} + \mathbb{T}_{\bowtie}(\mathbb{G}_{0}\mathbb{T}_{f} + \mathbb{I}))$ to find that the total extinct power, relative to a vacuum background, is given by
\begin{align}
    P^{\text{ext}}_{\mathrm{flx}} &=
    \frac{1}{2Z}\im{k_{0}\left<\mathbf{E}_{i}\right| \mathbb{T}_{\mathrm{tot}}\left|\mathbf{E}_{i}\right> } \\
    &= \frac{1}{2Z}\im{k_{0}\left<\mathbf{E}_{i}\right|\mathbb{W}_{f}\mathbb{V}_{f}\left|\mathbf{E}_{i}\right> + k_{0}\left<\mathbf{E}_{i}\right|\mathbb{W}_{f}\mathbb{T}_{\bowtie}(\mathbb{G}_{0}\mathbb{T}_{f} + \mathbb{I})\left|\mathbf{E}_{i}\right> } \\
    &= \frac{1}{2Z}\im{k_{0}\left<\mathbf{E}_{i}\right|\mathbb{W}_{f}\mathbb{V}_{f}\left|\mathbf{E}_{i}\right> }
    +
    \frac{1}{2Z}\im{k_{0}\left<\mathbf{E}_{i}\right|\mathbb{W}_{f}\left|\textbf{T}_{\bowtie}\right> }
    \label{eq:supp_clk_objDressed}
\end{align}
with $\mathbb{T}_{\bowtie}$ subject to the defining relation 
\begin{align}
  \mathbb{I}_{d} &= 
  \mathbb{I}_{d}\left(\chi_{d}^{-1}\mathbb{I} - \mathbb{G}_{f}\right)\mathbb{I}_{d}\mathbb{T}_{\bowtie}
  = 
  \mathbb{I}_{d}\left(\chi_{d}^{-1}\mathbb{I} - \left(\mathbb{G}_{0}+ \mathbb{G}_{0}\mathbb{T}_{f}\mathbb{G}_{0}\right)\right)\mathbb{I}_{d}\mathbb{T}_{\bowtie}
  =
  \mathbb{I}_{d}\left(\chi_{d}^{-1}\mathbb{I} - \mathbb{G}_{0}\mathbb{W}_{f}\right)\mathbb{I}_{d}\mathbb{T}_{\bowtie},
  \label{eq:supp_clk_tFundaDress}
\end{align}
where $\chi_{d}$ is the susceptibility used for the object in the design domain (the cloak).
The constraint is the same as in previous works~\cite{molesky_global_2020,chao_maximum_2022,strekha_suppressing_2024}, but where the relevant Green's function is $\mathbb{G}_{f}$ and not the vacuum Green's function.
Noting that $\mathbb{W}_{f} = \mathbb{G}_{0}^{-1}\mathbb{G}_{f},$ it is possible to rewrite the objective function in terms of $\mathbb{V}_{f}, \mathbb{V}_{d}, \mathbb{G}_{0}, \mathbb{G}_{f}$ (and/or their inverses),
\begin{align}
    P^{\text{ext}}_{\mathrm{flx}}
    &= \frac{1}{2Z}\im{k_{0}\left<\mathbf{E}_{i}\right|\mathbb{G}_{0}^{-1}\mathbb{G}_{f}\mathbb{V}_{f}\left|\mathbf{E}_{i}\right> }
    +
    \frac{1}{2Z}\im{k_{0}\left<\mathbf{E}_{i}\right|\mathbb{G}_{0}^{-1}\mathbb{G}_{f}\left|\textbf{T}_{\bowtie}\right>}.
    \label{eq:supp_clk_objDressed2}
\end{align}

\section{Optimization Setup}
With this background, we are ready to state our first optimization attempt. We want to minimize
\begin{align}
    P^{\text{ext}}_{\mathrm{flx}} &= \frac{1}{2Z}\im{k_{0}\left<\mathbf{E}_{i}\right|\mathbb{W}_{f}\mathbb{V}_{f}\left|\mathbf{E}_{i}\right>}
    +
    \frac{1}{2Z}\im{k_{0}\left<\mathbf{E}_{i}\right|\mathbb{W}_{f}\left|\textbf{T}_{\bowtie}\right> } \\
    &= \frac{1}{2Z}\im{k_{0}\left<\mathbf{E}_{i}\right|\mathbb{G}_{0}^{-1}\mathbb{G}_{f}\mathbb{V}_{f}\left|\mathbf{E}_{i}\right>}
    +
    \frac{1}{2Z}\im{k_{0}\left<\mathbf{E}_{i}\right|\mathbb{G}_{0}^{-1}\mathbb{G}_{f}\left|\textbf{T}_{\bowtie}\right>}
\end{align}
over all $|\mathbf{T}_{\bowtie}\rangle$ with support in the design region, subject to physics-inspired constraints.
Taking the adjoint of Eq.~\eqref{eq:supp_clk_tFundaDress}, multiplying by $\mathbb{T}_{\bowtie}$ from the right gives
\begin{equation}
    \mathbb{T}_{\bowtie} = \mathbb{T}_{\bowtie}^{\dagger}\mathbb{U}\mathbb{T}_{\bowtie},
    \label{eq:Tbowtieconstraintoperator}
\end{equation}
where 
\begin{align}
    \mathbb{U} &\equiv 
    \mathbb{I}_{d}(\chi_{d}^{-1\dagger}\mathbb{I} - \mathbb{G}_{f}^{\dagger})\mathbb{I}_{d} \\
    &=
    \mathbb{I}_{d}(\chi_{d}^{-1\dagger}\mathbb{I} -(\mathbb{G}_{0}+\mathbb{G}_{0}\mathbb{T}_{f}\mathbb{G}_{0})^{\dagger})\mathbb{I}_{d}.
\end{align}
Consider $\asym{\mathbb{U}}$.  Since $\mathbb{G}_{f}$ is a response function, its Asym part must be positive semidefinite, and since $\im{\chi_{d}} \geq 0$ for passive systems it follows that $\asym{\mathbb{U}}$ is also positive semidefinite.
We will partially enforce the constraints using Eq.~\eqref{eq:Tbowtieconstraintoperator}. We say partially because those equations hold as operators, but we will only require that equality hold for some bras and kets. We define
\begin{align}
    |\mathbf{E}_{\bowtie}\rangle &\equiv (\mathbb{G}_{0}\mathbb{T}_{f} + \mathbb{I})|\mathbf{E}_{i}\rangle = \mathbb{G}_{f}\mathbb{G}_{0}^{-1}|\mathbf{E}_{i}\rangle =  \frac{iZ}{k_{0}}\mathbb{G}_{f}\left|\mathbf{J}_{i}\right>, \\
    |\mathbf{T}_{\bowtie}\rangle &\equiv \mathbb{T}_{\bowtie}(\mathbb{G}_{0}\mathbb{T}_{f} + \mathbb{I})|\mathbf{E}_{i}\rangle = \mathbb{T}_{\bowtie}|\mathbf{E}_{\bowtie}\rangle.
\end{align}
Therefore, we multiply the constraints by 
$(\mathbb{G}_{0}\mathbb{T}_{f} +\mathbb{I})$ on the right and its adjoint on the left, and then sandwich it with $|\mathbf{E}_{i}\rangle,$ and take the real and imaginary parts of the resulting scalar equation. 
This leads to the optimization problem
\begin{subequations}
\begin{align}
    \underset{|\mathbf{T}_{\bowtie}\rangle}{\text{minimize}} \,\, 
    % \min_{|\mathbf{T}_{\bowtie}\rangle} \,\, 
P^{\text{ext}}_{\mathrm{flx}} &= \frac{1}{2Z}\im{k_{0}\left<\mathbf{E}_{i}\right|\mathbb{G}_{0}^{-1}\mathbb{G}_{f}\mathbb{V}_{f}\left|\mathbf{E}_{i}\right>}
+
\frac{1}{2Z}\im{k_{0}\left<\mathbf{E}_{i}\right|\mathbb{G}_{0}^{-1}\mathbb{G}_{f}\mathbb{I}_{d}\left|\textbf{T}_{\bowtie}\right>}
    \nonumber \\
\textrm{such that} \quad & \nonumber \\
    \im{\langle \mathbf{E}_{\bowtie}|\mathbb{I}_{d}|\mathbf{T}_{\bowtie}\rangle}
    &-\langle \mathbf{T}_{\bowtie}|\asym{\mathbb{U}}|\mathbf{T}_{\bowtie}\rangle = 0,
    \\
    \re{\langle \mathbf{E}_{\bowtie}|\mathbb{I}_{d}|\mathbf{T}_{\bowtie}\rangle}
    &-\langle \mathbf{T}_{\bowtie}|\sym{\mathbb{U}}|\mathbf{T}_{\bowtie}\rangle = 0,
\end{align}
\end{subequations}
where we inserted $\mathbb{I}_{d}$ in front of $|\mathbf{T}_{\bowtie}\rangle$ because we want the support to be within the design region ($\mathbb{T}_{\bowtie} = \mathbb{I}_{d}\mathbb{T}_{\bowtie}\mathbb{I}_{d}$).
% Said differently, we can take the optimization variable to be $|\mathbf{T}_{\bowtie}\rangle$ such that it has support over all space, and then the operators in the objective projects it back to the design region, or we state the minimization as $|\mathbf{T}_{\bowtie}\rangle$ such that supp$(|\mathbf{T}_{\bowtie}\rangle) \subseteq V_{d}$.

\subsection{Performance trick - Sparse formulation}

For a localized spatial basis (e.g., a finite difference grid), $\mathbb{G}_{f}$ is dense while $\mathbb{G}_{f}^{-1}$ is sparse. 
Define $\tilde{\mathbb{G}}_{f,dd} \equiv \mathbb{I}_{d}\mathbb{G}_{f}\mathbb{I}_{d}$.
In our notation, $\tilde{\mathbb{G}}_{f,dd}$ differs from $\mathbb{G}_{f,dd}$ in the domain and codomain (computationally, the matrix dimensions).
The size of $\tilde{\mathbb{G}}_{f,dd}$ is determined by the size of the computational grid, while $\mathbb{G}_{f,dd}$ is the subblock of the $\mathbb{G}_{f}$ operator for the spatial points in the design region.
That is,
\begin{align}
    \tilde{\mathbb{G}}_{f,dd}
    &=
    \begin{bmatrix}
        \mathbb{G}_{f,dd} & 0_{d\bar{d}} \\
        0_{\bar{d}d} & 0_{\bar{d}\bar{d}} 
    \end{bmatrix}
\end{align}
where $\bar{d}$ refers to the part of the domain that is not the design domain.
Also, in our notation $\tilde{\mathbb{G}}_{f,dd}\tilde{\mathbb{G}}_{f,dd}^{-1} = \mathbb{I}_{d}$, that is,
\begin{align}
    \tilde{\mathbb{G}}_{f,dd}^{-1}
    &=
    \begin{bmatrix}
        \mathbb{G}_{f,dd}^{-1} & 0_{d\bar{d}} \\
        0_{\bar{d}d} & 0_{\bar{d}\bar{d}} 
    \end{bmatrix}.
\end{align}
We find
\begin{align}
    \mathbb{I}_{d} &= 
    \mathbb{T}_{\bowtie}^{\dagger}\mathbb{I}_{d}\left(\chi_{d}^{-1\dagger}\mathbb{I} - \mathbb{G}_{f}^{\dagger}\right)\mathbb{I}_{d}, \\
    \mathbb{I}_{d}\mathbb{T}_{\bowtie} &= \mathbb{T}_{\bowtie}^{\dagger}\mathbb{I}_{d}(\chi_{d}^{-1\dagger}\mathbb{I} - \mathbb{G}_{f}^{\dagger})\mathbb{I}_{d}\mathbb{T}_{\bowtie}, \\
    \mathbb{I}_{d}\mathbb{T}_{\bowtie} &= \mathbb{T}_{\bowtie}^{\dagger}(\chi_{d}^{-1\dagger}\mathbb{I}_{d} - \tilde{\mathbb{G}}_{f,dd}^{\dagger})\mathbb{T}_{\bowtie}, \\
    \mathbb{I}_{d}\mathbb{T}_{\bowtie} &= \mathbb{T}_{\bowtie}^{\dagger}\tilde{\mathbb{G}}_{f,dd}^{\dagger}(\chi_{d}^{-1\dagger}\tilde{\mathbb{G}}_{f,dd}^{-1\dagger}\mathbb{I}_{d}\tilde{\mathbb{G}}_{f,dd}^{-1} - \tilde{\mathbb{G}}_{f,dd}^{-1})\tilde{\mathbb{G}}_{f,dd}\mathbb{T}_{\bowtie},
    \\
    \mathbb{I}_{d}\tilde{\mathbb{G}}_{f,dd}^{-1}\tilde{\mathbb{G}}_{f,dd}\mathbb{T}_{\bowtie} &= \mathbb{T}_{\bowtie}^{\dagger}\tilde{\mathbb{G}}_{f,dd}^{\dagger}(\chi_{d}^{-1\dagger}\tilde{\mathbb{G}}_{f,dd}^{-1\dagger}\mathbb{I}_{d}\tilde{\mathbb{G}}_{f,dd}^{-1} - \tilde{\mathbb{G}}_{f,dd}^{-1})\tilde{\mathbb{G}}_{f,dd}\mathbb{T}_{\bowtie},
\end{align}
and so, defining 
\begin{align}
    |\mathbf{E}_{\bowtie}\rangle &\equiv (\mathbb{G}_{0}\mathbb{T}_{f} + \mathbb{I})|\mathbf{E}_{i}\rangle = \mathbb{G}_{f}\mathbb{G}_{0}^{-1}|\mathbf{E}_{i}\rangle, \\
    |\mathbf{T}_{\bowtie}\rangle &\equiv \mathbb{T}_{\bowtie}(\mathbb{G}_{0}\mathbb{T}_{f} + \mathbb{I})|\mathbf{E}_{i}\rangle = \mathbb{T}_{\bowtie}|\mathbf{E}_{\bowtie}\rangle, \\
|\tilde{\textbf{T}}_{\bowtie}\rangle &\equiv \tilde{\mathbb{G}}_{f,dd}\left|\textbf{T}_{\bowtie}\right>, \\
    \tilde{\mathbb{U}} &\equiv
    \mathbb{I}_{d}(\chi_{d}^{-1\dagger}\tilde{\mathbb{G}}_{f,dd}^{-1\dagger}\mathbb{I}_{d}\tilde{\mathbb{G}}_{f,dd}^{-1} - \tilde{\mathbb{G}}_{f,dd}^{-1})\mathbb{I}_{d},
\end{align}
one finds the constraints
\begin{align}
    \im{\langle \mathbf{E}_{\bowtie}|\mathbb{I}_{d}\tilde{\mathbb{G}}_{f,dd}^{-1}|\tilde{\mathbf{T}}_{\bowtie}\rangle}
    &=
    \langle \tilde{\mathbf{T}}_{\bowtie}|\asym{\tilde{\mathbb{U}}}|\tilde{\mathbf{T}}_{\bowtie}\rangle, \\
    \re{\langle \mathbf{E}_{\bowtie}|\mathbb{I}_{d}\tilde{\mathbb{G}}_{f,dd}^{-1}|\tilde{\mathbf{T}}_{\bowtie}\rangle}
    &=
    \langle \tilde{\mathbf{T}}_{\bowtie}|\sym{\tilde{\mathbb{U}}}|\tilde{\mathbf{T}}_{\bowtie}\rangle.
\end{align}

The optimization problem is thus
\begin{subequations}
\begin{align}
    \underset{|\tilde{\mathbf{T}}_{\bowtie}\rangle}{\text{minimize}} \,\, 
    % \min_{|\tilde{\mathbf{T}}_{\bowtie}\rangle} \,\, 
    P^{\text{ext}}_{\mathrm{flx}} &= \frac{1}{2Z}\im{k_{0}\left<\mathbf{E}_{i}\right|\mathbb{G}_{0}^{-1}\mathbb{G}_{f}\mathbb{V}_{f}\left|\mathbf{E}_{i}\right>}
    +
    \frac{1}{2Z}\im{k_{0}\left<\mathbf{E}_{i}\right|\mathbb{G}_{0}^{-1}\mathbb{G}_{f}\tilde{\mathbb{G}}_{f,dd}^{-1}|\tilde{\textbf{T}}_{\bowtie}\rangle}
    \nonumber \\
    % \mathrm{s.t.} \quad & \forall \vb{r} \in V \nonumber \\
    \textrm{such that} \quad & \nonumber \\
    &\im{\langle \mathbf{E}_{\bowtie}|\mathbb{I}_{d}\tilde{\mathbb{G}}_{f,dd}^{-1}|\tilde{\mathbf{T}}_{\bowtie}\rangle}
    -
    \langle \tilde{\mathbf{T}}_{\bowtie}|\asym{\tilde{\mathbb{U}}}|\tilde{\mathbf{T}}_{\bowtie}\rangle = 0, \\
    &\re{\langle \mathbf{E}_{\bowtie}|\mathbb{I}_{d}\tilde{\mathbb{G}}_{f,dd}^{-1}|\tilde{\mathbf{T}}_{\bowtie}\rangle}
    -
    \langle \tilde{\mathbf{T}}_{\bowtie}|\sym{\tilde{\mathbb{U}}}|\tilde{\mathbf{T}}_{\bowtie}\rangle = 0.
\end{align}
\end{subequations}
Thus, one needs to evaluate the fixed vectors
\begin{align}
    |\mathbf{V}_{1}\rangle &\equiv \mathbb{G}_{0}^{-1}\mathbb{G}_{f}\mathbb{V}_{f}|\mathbf{E}_{i}\rangle, \\
    |\mathbf{V}_{3}\rangle &\equiv \tilde{\mathbb{G}}_{f,dd}^{-1}\mathbb{G}_{f}^{}\mathbb{G}_{0}^{-1}|\mathbf{E}_{i}^{*}\rangle, \\
    |\mathbf{V}_{2}\rangle &\equiv \tilde{\mathbb{G}}_{f,dd}^{-1\dagger}\mathbb{G}_{f}^{\dagger}\mathbb{G}_{0}^{-1\dagger}|\mathbf{E}_{i}\rangle = |\mathbf{V}_{3}^{*}\rangle,
\end{align}
where the last line follows for reciprocal systems ($\mathbb{G}_{f} = \mathbb{G}_{f}^{T}$ and $\mathbb{G}_{0} = \mathbb{G}_{0}^{T}$).
Once these vectors are calculated, the term
\begin{align}
    \frac{1}{2Z}\im{k_{0}\left<\mathbf{E}_{i}\right|\mathbb{G}_{0}^{-1}\mathbb{G}_{f}\mathbb{V}_{f}\left|\mathbf{E}_{i}\right>}
    &=
    \frac{1}{2Z}\im{k_{0}\langle\mathbf{E}_{i}|\mathbf{V}_{1}\rangle}
\end{align}
is known and is a constant in the objective function. Also, in
\begin{align}
    \frac{1}{2Z}\im{k_{0}\left<\mathbf{E}_{i}\right|\mathbb{G}_{0}^{-1}\mathbb{G}_{f}\tilde{\mathbb{G}}_{f,dd}^{-1}\left|\tilde{\textbf{T}}_{\bowtie}\right>}
    &=
    \frac{1}{2Z}\im{k_{0}\langle\mathbf{V}_{2}
    |\tilde{\textbf{T}}_{\bowtie}\rangle}
\end{align}
the bra is known and the sparse formulation of the optimization problem over $|\tilde{\textbf{T}}_{\bowtie}\rangle$ is as follows:
\begin{subequations}
\begin{align}
    \underset{|\tilde{\mathbf{T}}_{\bowtie}\rangle}{\text{minimize}} \,\, 
    P^{\text{ext}}_{\mathrm{flx}} &= \frac{1}{2Z}\im{k_{0}\langle\mathbf{E}_{i}|\mathbf{V}_{1}\rangle}
    +
    \frac{1}{2Z}\im{k_{0}\langle\mathbf{V}_{2}
    |\tilde{\textbf{T}}_{\bowtie}\rangle}
    \nonumber \\
    \textrm{such that} \quad & \nonumber \\
    &\im{\langle \mathbf{E}_{\bowtie}|\mathbb{I}_{d}\tilde{\mathbb{G}}_{f,dd}^{-1}|\tilde{\mathbf{T}}_{\bowtie}\rangle}
    -
    \langle \tilde{\mathbf{T}}_{\bowtie}|\asym{\tilde{\mathbb{U}}}|\tilde{\mathbf{T}}_{\bowtie}\rangle = 0, \\
    &\re{\langle \mathbf{E}_{\bowtie}|\mathbb{I}_{d}\tilde{\mathbb{G}}_{f,dd}^{-1}|\tilde{\mathbf{T}}_{\bowtie}\rangle}
    -
    \langle \tilde{\mathbf{T}}_{\bowtie}|\sym{\tilde{\mathbb{U}}}|\tilde{\mathbf{T}}_{\bowtie}\rangle = 0.
\end{align}
\end{subequations}
Note also that the only nonzero parts of $|\mathbf{V}_{2}\rangle$ and $|\tilde{\textbf{T}}_{\bowtie}\rangle$ are those over the design region.
Thus, computationally we do not need the representation of the vectors and matrices defined over the entire computational grid, but only their restriction to the design region.
Likewise, we only need $\mathbb{I}_{d}|\mathbf{E}_{\bowtie}\rangle$ and not $|\mathbf{E}_{\bowtie}\rangle$. ($|\mathbf{V}_{1}\rangle$ can be nonzero outside the design region, but we calculate $\langle\mathbf{E}_{i}|\mathbf{V}_{1}\rangle$ separately and only once.)
Also, note that the constraints can be generalized with the additional application of an arbitrary projection operator $\mathbb{P}$ that commutes with $\mathbb{I}_{d}$ to start with
\begin{align}
    \mathbb{I}_{d}\mathbb{P} &= 
    \mathbb{T}_{\bowtie}^{\dagger}\mathbb{I}_{d}\left(\chi_{d}^{-1\dagger}\mathbb{I}\mathbb{P} - \mathbb{G}_{f}^{\dagger}\mathbb{P}\right)\mathbb{I}_{d} 
\end{align}
which ultimately leads to 
\begin{align}
    \mathbb{I}_{d}\mathbb{P}\tilde{\mathbb{G}}_{f,dd}^{-1}\tilde{\mathbb{G}}_{f,dd}\mathbb{T}_{\bowtie} &= \mathbb{T}_{\bowtie}^{\dagger}\tilde{\mathbb{G}}_{f,dd}^{\dagger}(\chi_{d}^{-1\dagger}\tilde{\mathbb{G}}_{f,dd}^{-1\dagger}\mathbb{I}_{d}\mathbb{P}\tilde{\mathbb{G}}_{f,dd}^{-1} - \mathbb{I}_{d}\mathbb{P}\tilde{\mathbb{G}}_{f,dd}^{-1})\tilde{\mathbb{G}}_{f,dd}\mathbb{T}_{\bowtie}
\end{align}
and the constraints
\begin{align}
    \im{\langle \mathbf{E}_{\bowtie}|\mathbb{I}_{d}\mathbb{P}\tilde{\mathbb{G}}_{f,dd}^{-1}|\tilde{\mathbf{T}}_{\bowtie}\rangle}
    -
    \langle \tilde{\mathbf{T}}_{\bowtie}|
    \asym{
    \chi_{d}^{-1\dagger}\tilde{\mathbb{G}}_{f,dd}^{-1\dagger}\mathbb{I}_{d}\mathbb{P}\tilde{\mathbb{G}}_{f,dd}^{-1} - \mathbb{I}_{d}\mathbb{P}\tilde{\mathbb{G}}_{f,dd}^{-1}
    }
    |\tilde{\mathbf{T}}_{\bowtie}\rangle &= 0, \\
    \re{\langle \mathbf{E}_{\bowtie}|\mathbb{I}_{d}\mathbb{P}\tilde{\mathbb{G}}_{f,dd}^{-1}|\tilde{\mathbf{T}}_{\bowtie}\rangle}
    -
    \langle \tilde{\mathbf{T}}_{\bowtie}|
    \sym{
    \chi_{d}^{-1\dagger}\tilde{\mathbb{G}}_{f,dd}^{-1\dagger}\mathbb{I}_{d}\mathbb{P}\tilde{\mathbb{G}}_{f,dd}^{-1} - \mathbb{I}_{d}\mathbb{P}\tilde{\mathbb{G}}_{f,dd}^{-1}
    }|\tilde{\mathbf{T}}_{\bowtie}\rangle &= 0.
\end{align}

\subsection{Final sparse optimization problem}

Separating the derivation from the main result, we write our final sparse formulation of the optimization problem:

\begin{subequations}
\begin{align}
    \underset{|\tilde{\mathbf{T}}_{\bowtie}\rangle}{\text{minimize}} \,\, 
    P^{\text{ext}}_{\mathrm{flx}} &= \frac{1}{2Z}\im{k_{0}\langle\mathbf{E}_{i}|\mathbf{V}_{1}\rangle}
    +
    \frac{1}{2Z}\im{k_{0}\langle\mathbf{V}_{2}
    |\tilde{\textbf{T}}_{\bowtie}\rangle}
    \nonumber \\
% \mathrm{s.t.} \quad & \forall \vb{r} \in V \nonumber \\
\textrm{such that} \quad & \nonumber \\
    &\im{\langle \mathbf{E}_{\bowtie}|\mathbb{I}_{d}\mathbb{P}\tilde{\mathbb{G}}_{f,dd}^{-1}|\tilde{\mathbf{T}}_{\bowtie}\rangle}
    -
    \langle \tilde{\mathbf{T}}_{\bowtie}|
    \asym{
    \chi_{d}^{-1\dagger}\tilde{\mathbb{G}}_{f,dd}^{-1\dagger}\mathbb{I}_{d}\mathbb{P}\tilde{\mathbb{G}}_{f,dd}^{-1} - \mathbb{I}_{d}\mathbb{P}\tilde{\mathbb{G}}_{f,dd}^{-1}
    }
    |\tilde{\mathbf{T}}_{\bowtie}\rangle = 0, \\
    &\re{\langle \mathbf{E}_{\bowtie}|\mathbb{I}_{d}\mathbb{P}\tilde{\mathbb{G}}_{f,dd}^{-1}|\tilde{\mathbf{T}}_{\bowtie}\rangle}
    -
    \langle \tilde{\mathbf{T}}_{\bowtie}|
    \sym{
    \chi_{d}^{-1\dagger}\tilde{\mathbb{G}}_{f,dd}^{-1\dagger}\mathbb{I}_{d}\mathbb{P}\tilde{\mathbb{G}}_{f,dd}^{-1} - \mathbb{I}_{d}\mathbb{P}\tilde{\mathbb{G}}_{f,dd}^{-1}
    }|\tilde{\mathbf{T}}_{\bowtie}\rangle = 0,
\end{align}
\end{subequations}
where
\begin{align}
    |\mathbf{V}_{1}\rangle &\equiv \mathbb{G}_{0}^{-1}\mathbb{G}_{f}\mathbb{V}_{f}|\mathbf{E}_{i}\rangle, \\
    |\mathbf{V}_{3}\rangle &\equiv \tilde{\mathbb{G}}_{f,dd}^{-1}\mathbb{G}_{f}^{}\mathbb{G}_{0}^{-1}|\mathbf{E}_{i}^{*}\rangle, \\
    |\mathbf{V}_{2}\rangle &\equiv \tilde{\mathbb{G}}_{f,dd}^{-1\dagger}\mathbb{G}_{f}^{\dagger}\mathbb{G}_{0}^{-1\dagger}|\mathbf{E}_{i}\rangle = |\mathbf{V}_{3}^{*}\rangle, \\
    |\mathbf{E}_{\bowtie}\rangle &\equiv (\mathbb{G}_{0}\mathbb{T}_{f} + \mathbb{I})|\mathbf{E}_{i}\rangle = \mathbb{G}_{f}\mathbb{G}_{0}^{-1}|\mathbf{E}_{i}\rangle =  \frac{iZ}{k_{0}}\mathbb{G}_{f}\left|\mathbf{J}_{i}\right>.
    % \\
    % \mathbb{I}_{d}|\mathbf{E}_{\bowtie}\rangle &= \mathbb{I}_{d}\mathbb{G}_{f}\mathbb{G}_{0}^{-1}|\mathbf{E}_{i}\rangle =  \frac{iZ}{k_{0}}\mathbb{I}_{d}\mathbb{G}_{f}\left|\mathbf{J}_{i}\right>.
\end{align}
As a note of caution, $\tilde{\mathbb{G}}_{f,dd}^{-1}$ means the inverse of the projection of $\tilde{\mathbb{G}}_{f}$ onto design domain, and not the inverse of $\mathbb{G}_{f}$ which then gets projected onto the design domain.
That is, $\tilde{\mathbb{G}}_{f,dd}^{-1} = (\tilde{\mathbb{G}}_{f,dd})^{-1}$ and not $\mathbb{I}_{d}(\mathbb{G}_{f}^{-1})\mathbb{I}_{d}.$

\section{Inverse design procedure}

Let $\{\bar{\chi}_{k}\}$ be the topology optimization degrees of freedom, normalized to take values in $[0, 1]$ where $k$ indexes a pixel of the computational grid. This can be interpreted as a filling fraction of the pixel with the design material.
Let $\chi_{\mathrm{des}}$ be the value of the electric susceptibility used in the design region (the cloak). Then the electric susceptibility in the design region part of the computational grid is given by $\chi_{k} = \chi_{\mathrm{des}}\bar{\chi}_{k}$ and the objective function can be viewed as a function of $\{\bar{\chi}_{k}\}$,
\begin{align}
    P_{\mathrm{flx}}^{\text{ext}}(\{\bar{\chi}_{k}\};\omega_{0}) =
    \frac{1}{2}\re{\left<\mathbf{E}_{i}|\mathbf{J}_{g}\right>} = 
  \frac{1}{2Z}\im{k_{0}\left<\mathbf{E}_{i}|\mathbb{V}|\mathbf{E}_{t}\right>} =
  \frac{1}{2}\re{\left<\mathbf{E}_{i}|\mathbb{V}\mathbb{G}|\mathbf{J}_{i}\right>},
\end{align}
where $\mathbb{G}$ is the total Green's function. 
Let $\mathbb{M}^{-1} = \mathbb{G}$ where $\mathbb{M} = \frac{c^{2}}{\omega_{0}^{2}}\nabla\times\nabla - \epsilon(\mathbf{r})$.
After the discretization of the computational grid, then in the design region $\epsilon_{k} = 1 + \chi_{\mathrm{des}}\bar{\chi}_{k}$ at pixel $k$. It follows that
\begin{align}
    \frac{\partial P_{\mathrm{flx}}^{\text{ext}}(\{\bar{\chi}_{k}\};\omega_{0})}{\partial \bar{\chi}_{a}}
    &=
    \frac{1}{2}\re{\left<\mathbf{E}_{i}|\frac{\partial(\mathbb{V}\mathbb{G})}{\partial \bar{\chi}_{a}}|\mathbf{J}_{i}\right>}
    \\
    &=
    \frac{1}{2}\re{\left<\mathbf{E}_{i}|
    (\chi_{\mathrm{des}}\mathbb{P}_{a}\mathbb{G} + \mathbb{V} \frac{\partial \mathbb{M}^{-1}}{\partial \bar{\chi}_{a}})
    |\mathbf{J}_{i}\right>} \\
    &=
    \frac{1}{2}\re{\left<\mathbf{E}_{i}|
    (\chi_{\mathrm{des}}\mathbb{P}_{a}\mathbb{G} - \mathbb{V} \mathbb{M}^{-1}\frac{\partial \mathbb{M}}{\partial \bar{\chi}_{a}}\mathbb{M}^{-1})
    |\mathbf{J}_{i}\right>} \\
    &=
    \frac{1}{2}\re{\left<\mathbf{E}_{i}|
    (\chi_{\mathrm{des}}\mathbb{P}_{a}\mathbb{G} 
    + \chi_{\mathrm{des}}\mathbb{V} \mathbb{G}\mathbb{P}_{a}\mathbb{G})
    |\mathbf{J}_{i}\right>} \\
    &=
    \frac{1}{2}\re{\frac{k_{0}\chi_{\mathrm{des}}}{iZ}\left<\mathbf{E}_{i}|
    \mathbb{P}_{a} |\mathbf{E}_{t}\right>}
    + 
    \frac{1}{2}\re{\frac{k_{0}\chi_{\mathrm{des}}}{iZ}\left<\mathbf{E}_{i}|\mathbb{V} \mathbb{G}\mathbb{P}_{a}
    |\mathbf{E}_{t}\right>} \\
    &\equiv
    \frac{1}{2}\re{\frac{k_{0}\chi_{\mathrm{des}}}{iZ}\left<\mathbf{E}_{i}|
    \mathbb{P}_{a} |\mathbf{E}_{t}\right>}
    + 
    \frac{1}{2}\re{\frac{k_{0}\chi_{\mathrm{des}}}{iZ} \langle\tilde{\mathbf{E}}|\mathbb{P}_{a}
    |\mathbf{E}_{t}\rangle}
\end{align}
where $\mathbb{P}_{a}$ is projection onto the pixel indexed by $a$, and $|\tilde{\mathbf{E}}\rangle \equiv \mathbb{G}^{\dagger}\mathbb{V}^{\dagger}|\mathbf{E}_{i}\rangle$ which for reciprocal systems becomes $|\tilde{\mathbf{E}}\rangle \equiv (\mathbb{G}\mathbb{V}|\mathbf{E}_{i}^{*}\rangle)^{*}$.
Thus, for a given incident field $|\mathbf{E}_{i}\rangle$ one only needs to perform two Maxwell solves to compute $|\mathbf{E}^{t}\rangle$ and $|\tilde{\mathbf{E}}\rangle$ at each iteration of the optimization to obtain the required gradient. 
Furthermore, note that no additional Maxwell solves are needed to calculate the objective value since 
$P_{\mathrm{flx}}^{\text{ext}}(\{\bar{\chi}_{k}\};\omega_{0}) = \frac{1}{2Z}\im{k_{0}\left<\mathbf{E}_{i}|\mathbb{V}|\mathbf{E}_{t}\right>}$.

\section{Deriving asymptotics from T operator expressions}

First, write the operators in block-form for two regions, labeled A and B.
\begin{align}
    \mathbb{G}^{0} &= 
    \begin{bmatrix}
        \mathbb{G}^{0}_{AA} & \mathbb{G}^{0}_{AB} \\
        \mathbb{G}^{0}_{BA} & \mathbb{G}^{0}_{BB}
    \end{bmatrix}, \\
    \mathbb{V}^{-1} &=
    \begin{bmatrix}
        \mathbb{V}^{-1}_{A} & 0 \\
        0 & \mathbb{V}^{-1}_{B}
    \end{bmatrix},
\end{align}
which then leads to the total $\mathbb{T}$ operator
\begin{align}
    \mathbb{T}^{-1} &=
    \begin{bmatrix}
        \mathbb{T}_{A}^{-1} & -\mathbb{G}^{0}_{AB} \\
        -\mathbb{G}^{0}_{BA} & \mathbb{T}_{B}^{-1}
    \end{bmatrix}, \\
    \mathbb{T} &=
    \begin{bmatrix}
        (\mathbb{T}_{A}^{-1} - \mathbb{G}^{0}_{AB}\mathbb{T}_{B}\mathbb{G}^{0}_{BA})^{-1} & \mathbb{T}_{A}\mathbb{G}^{0}_{AB}(\mathbb{T}_{B}^{-1} - \mathbb{G}^{0}_{BA}\mathbb{T}_{A}\mathbb{G}^{0}_{AB})^{-1} \\
        \mathbb{T}_{B}\mathbb{G}^{0}_{BA}(\mathbb{T}_{A}^{-1} - \mathbb{G}^{0}_{AB}\mathbb{T}_{B}\mathbb{G}^{0}_{BA})^{-1} & (\mathbb{T}_{B}^{-1} - \mathbb{G}^{0}_{BA}\mathbb{T}_{A}\mathbb{G}^{0}_{AB})^{-1}
    \end{bmatrix} \\
    &\equiv 
    \begin{bmatrix}
        \mathbb{T}_{AA} & \mathbb{T}_{AB} \\
        \mathbb{T}_{BA} & \mathbb{T}_{BB}
    \end{bmatrix},
\end{align}
where
\begin{align}
    \mathbb{T}_{AA} &\equiv (\mathbb{T}_{A}^{-1} - \mathbb{G}^{0}_{AB}\mathbb{T}_{B}\mathbb{G}^{0}_{BA})^{-1} \equiv \mathbb{Y}_{A}, \\
    \mathbb{T}_{AB} &\equiv \mathbb{T}_{A}\mathbb{G}^{0}_{AB}(\mathbb{T}_{B}^{-1} - \mathbb{G}^{0}_{BA}\mathbb{T}_{A}\mathbb{G}^{0}_{AB})^{-1} = \mathbb{T}_{A}\mathbb{G}^{0}_{AB}\mathbb{Y}_{B}, \\
    \mathbb{T}_{BA} &\equiv \mathbb{T}_{B}\mathbb{G}^{0}_{BA}(\mathbb{T}_{A}^{-1} - \mathbb{G}^{0}_{AB}\mathbb{T}_{B}\mathbb{G}^{0}_{BA})^{-1} = \mathbb{T}_{B}\mathbb{G}^{0}_{BA}\mathbb{Y}_{A}, \\
    \mathbb{T}_{BB} &\equiv (\mathbb{T}_{B}^{-1} - \mathbb{G}^{0}_{BA}\mathbb{T}_{A}\mathbb{G}^{0}_{AB})^{-1} \equiv \mathbb{Y}_{B}.
\end{align}
Note that $\mathbb{Y}_{A}$ is the scattering operator of object $A$ dressed by the presence of object $B$. Namely,
\begin{align}
    \mathbb{Y}_{A} &= (\mathbb{T}_{A}^{-1} - \mathbb{G}^{0}_{AB}\mathbb{T}_{B}\mathbb{G}^{0}_{BA})^{-1} \\
    &= (\mathbb{V}_{A}^{-1} - \mathbb{G}^{0}_{AA} - \mathbb{G}^{0}_{AB}\mathbb{T}_{B}\mathbb{G}^{0}_{BA})^{-1} \\
    &= (\mathbb{V}_{A}^{-1} - \mathbb{I}_{A}(\mathbb{G}^{0} + \mathbb{G}^{0}\mathbb{T}_{B}\mathbb{G}^{0})\mathbb{I}_{A})^{-1} \\
    &= (\mathbb{V}_{A}^{-1} -\mathbb{G}^{B}_{AA})^{-1}
\end{align}
where $\mathbb{G}^{B} = \mathbb{G}^{0} + \mathbb{G}^{0}\mathbb{T}_{B}\mathbb{G}^{0}$ is the Green's function of object $B$ in isolation. The same holds for $\mathbb{Y}_{B}$ with $A$ and $B$ changing roles.
Without loss of generality, focus on $\mathbb{V}_{B} = (\chi_{B}' + i\chi_{B}'')\mathbb{I}_{B}$ with $\chi_{B}'' \to 0$. Starting from
\begin{align}
    \mathbb{T}_{B} &= \mathbb{V}_{B} (\mathbb{I} - \mathbb{G}^{0}_{BB}\mathbb{V}_{B})^{-1} \\
    &= (\chi_{B}'\mathbb{I}_{B})(\mathbb{I} - \mathbb{G}^{0}_{BB}\mathbb{V}_{B})^{-1} + (i\chi_{B}''\mathbb{I}_{B})(\mathbb{I} - \mathbb{G}^{0}_{BB}\mathbb{V}_{B})^{-1}
\end{align}
we expand in terms of $\chi_{B}''$ to get
\begin{align}
    \mathbb{T}_{B}
    = 
    (\chi_{B}'\mathbb{I}_{B})[(\mathbb{I} - \mathbb{G}^{0}_{BB}\mathbb{V}_{B}')^{-1} + \sum_{n=0}^{\infty}{n \choose 1}(\mathbb{G}^{0}_{BB})^{n}(\chi_{B}')^{n-1}(i\chi_{B}'')] + (i\chi_{B}''\mathbb{I}_{B})(\mathbb{I} - \mathbb{G}^{0}_{BB}\mathbb{V}_{B}')^{-1} + O[(\chi_{B}'')^{2}].
\end{align}

\section{Additional figures of the performance of inverse designs}

\begin{figure}
    \centering
        \includegraphics[width=0.60\linewidth]{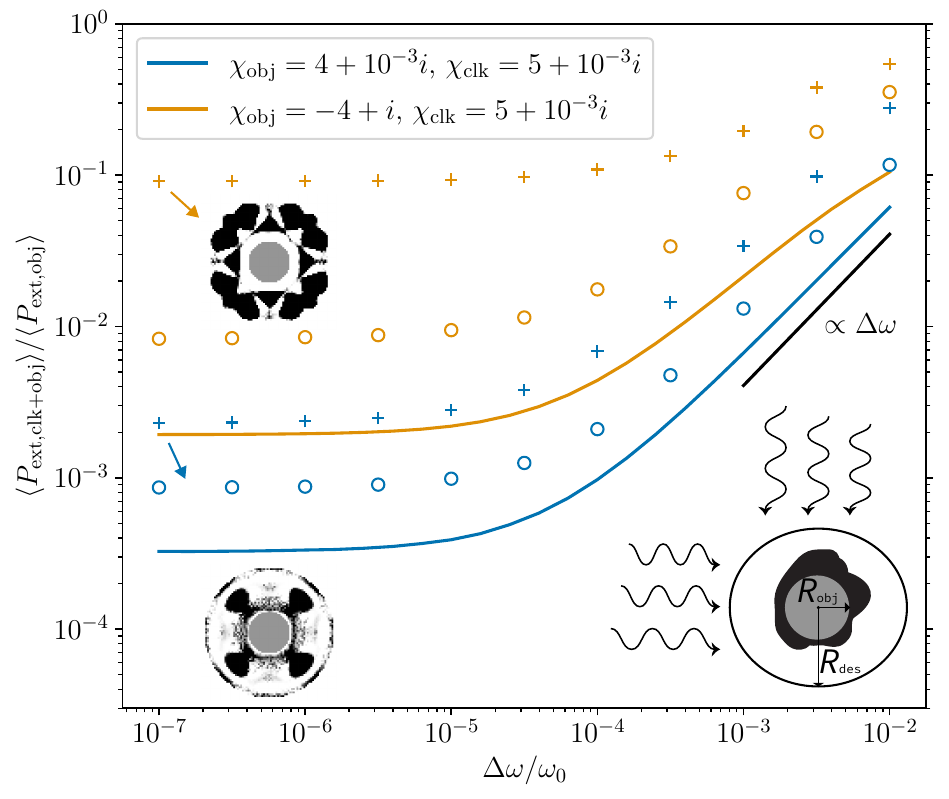}
        \caption{
        Bounds on bandwidth-integrated extinction as a function of the bandwidth, with the same parameters as in Figure 1 of the main text but where the pluses correspond to inverse designs for a figure of merit with a weighted sum of incident vertical and horizontal plane waves (bottom-right schematic).
        }
\label{fig:bounds_vs_bandwidth_2sides_supp}
\end{figure}

\begin{figure}
    \centering
        \includegraphics[width=0.75\linewidth]{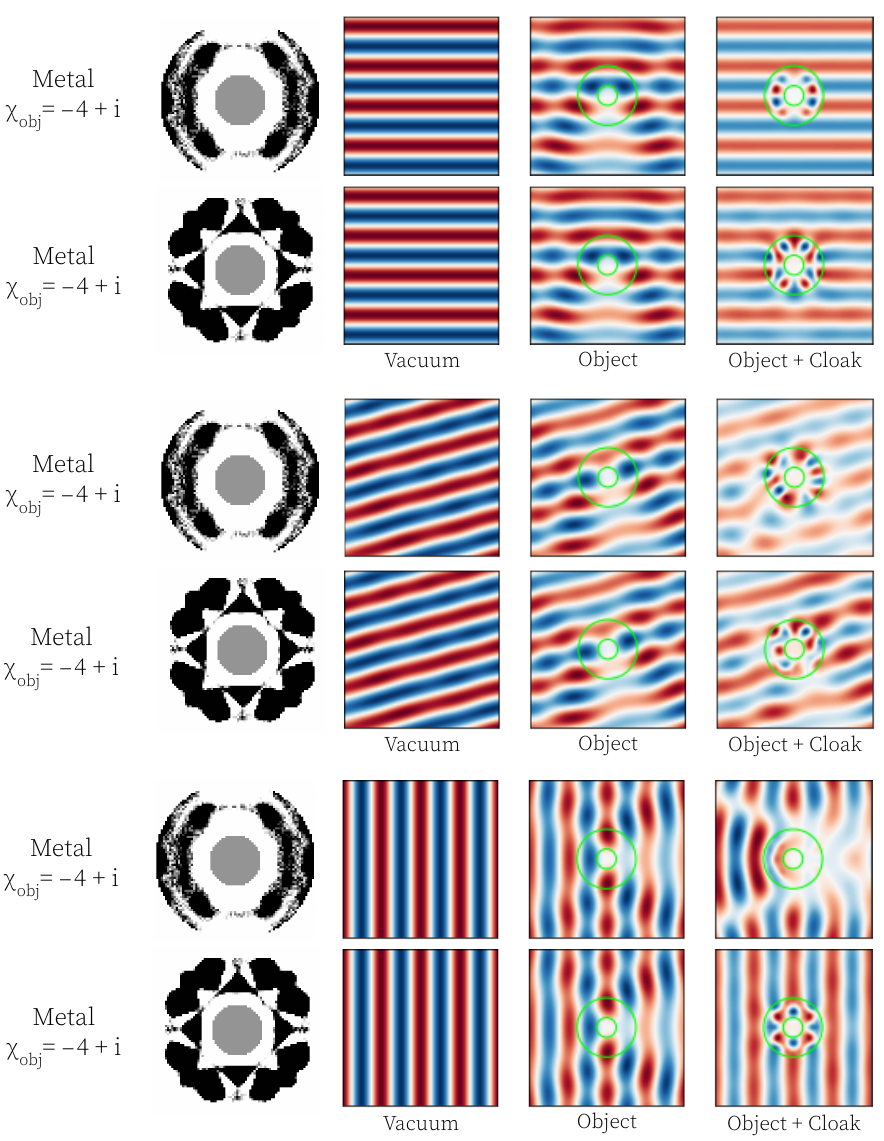}
        \caption{
        (Left) Cloaking of a metal object with $\chi_{\obj} = -4 + i$ by a cloak with susceptibility $\chi_{\clk} = 5 + 10^{-3}i$. (Right) Shown are the real parts of the out-of-plane electric fields. The structure optimized for two orthogonal angles of incidence might perform a bit worse for the vertical incident plane wave (top two rows) but performs better for intermediate angles and horizontal plane waves (bottom four rows).
        }
\label{fig:metal_1side_2side_comparison}
\end{figure}

\begin{figure}
    \centering
        \includegraphics[width=0.75\linewidth]{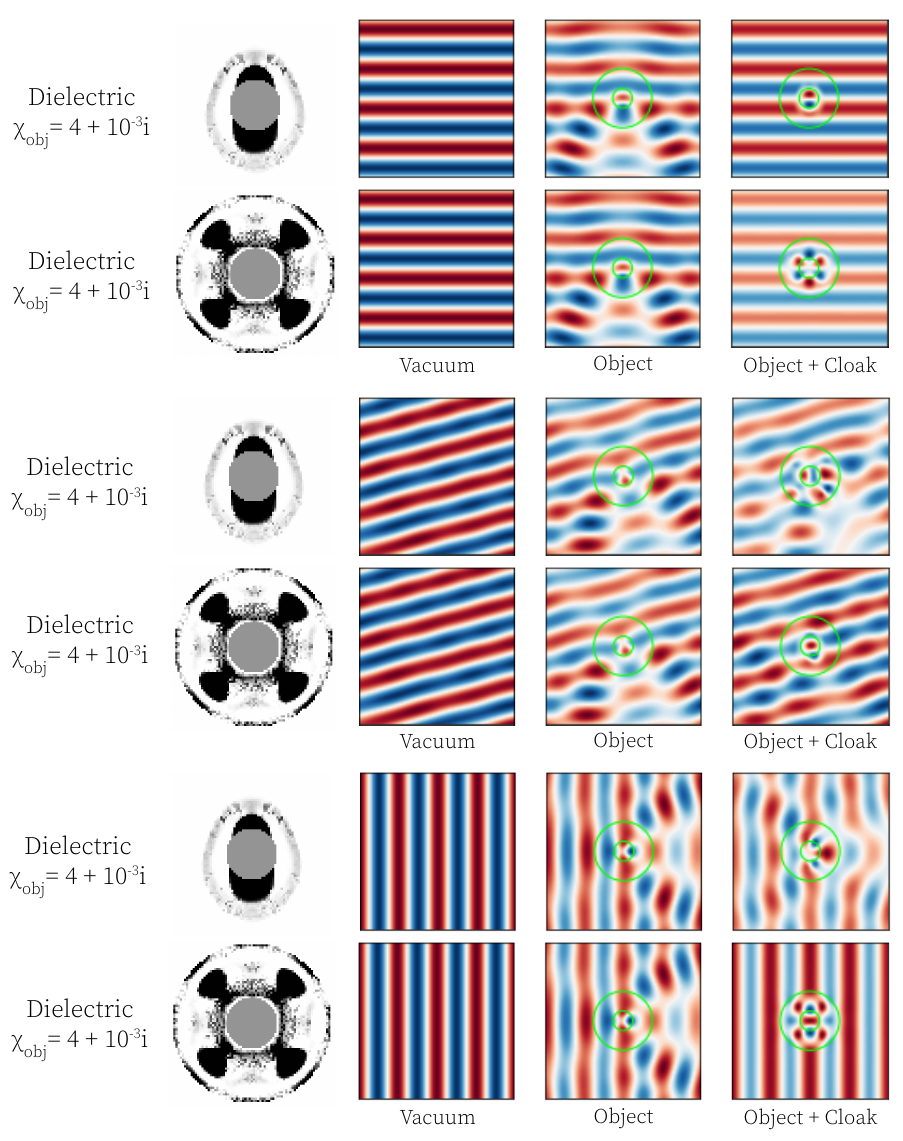}
        \caption{
        (Left) Cloaking of a dielectric object with $\chi_{\obj} = 4 + 10^{-3}i$ by a cloak with susceptibility $\chi_{\clk} = 5 + 10^{-3}i$.
        (Right) Shown are the real parts of the out-of-plane electric fields.
        The structure optimized for two orthogonal angles of incidence displays better cloaking performance for intermediate angles and horizontal plane waves (bottom four rows).
        }
\label{fig:dielectric_1side_2side_comparison}
\end{figure}

Shown in Fig.~\ref{fig:metal_1side_2side_comparison} and Fig.~\ref{fig:dielectric_1side_2side_comparison} are the real parts of the out-of-plane electric fields in the presence and absence of the cloak for structures discovered via structural optimization with an objective function with an incident plane from one direction and two orthogonal directions. Figure~\ref{fig:bounds_vs_bandwidth_2sides_supp} shows the performance values of such devices in comparison to those of devices optimized for only one direction of incident plane waves.
That is, we consider two incidence directions $\phi_{1}$ and $\phi_{2}$ that satisfy $|\phi_{2} - \phi_{1}| = \pi/2$ with an optimization problem of the form
\begin{subequations}
\begin{align}
\underset{|\textbf{T}_{\bowtie}^{(1)}\rangle, |\textbf{T}_{\bowtie}^{(2)}\rangle}{\text{minimize}} 
\quad &  \frac{1}{2Z}\im{k_{0}\left<\mathbf{E}_{i}^{(1)}\right|\mathbb{G}_{0}^{-1}\mathbb{G}_{f}\mathbb{V}_{f}\left|\mathbf{E}_{i}^{(1)}\right>} 
    + \frac{1}{2Z}\im{k_{0}\left<\mathbf{E}_{i}^{(1)}\right|\mathbb{G}_{0}^{-1}\mathbb{G}_{f}\mathbb{P}_{\textrm{des}}\left|\textbf{T}_{\bowtie}^{(1)}\right>} \nonumber
    \\
     + &\frac{1}{2Z}\im{k_{0}\left<\mathbf{E}_{i}^{(2)}\right|\mathbb{G}_{0}^{-1}\mathbb{G}_{f}\mathbb{V}_{f}\left|\mathbf{E}_{i}^{(2)}\right>} 
    + \frac{1}{2Z}\im{k_{0}\left<\mathbf{E}_{i}^{(2)}\right|\mathbb{G}_{0}^{-1}\mathbb{G}_{f}\mathbb{P}_{\textrm{des}}\left|\textbf{T}_{\bowtie}^{(2)}\right>} \label{eq:cloakingPrimalProblemTwoSides} \\
\text{such that } \forall k & \nonumber \\
\quad & \im{\langle \mathbf{E}_{\bowtie}^{(a)}|\mathbb{P}_{k}|\mathbf{T}_{\bowtie}^{(b)}\rangle
    -\langle \mathbf{T}_{\bowtie}^{(a)}|\mathbb{U}_{k}|\mathbf{T}_{\bowtie}^{(b)}\rangle} = 0,
    \\
    &\re{\langle \mathbf{E}_{\bowtie}^{(a)}|\mathbb{P}_{k}|\mathbf{T}_{\bowtie}^{(b)}\rangle
    -\langle \mathbf{T}_{\bowtie}^{(a)}|\mathbb{U}_{k}|\mathbf{T}_{\bowtie}^{(b)}\rangle} = 0,
\end{align}
\end{subequations}
for $a,b \in \{1, 2\}$. Without the constraints between the $a=1, b=2$ and $a=2, b=1$ terms, the problem reduces to a sum of decoupled optimization problems.

\section{Calculation of the Lagrange dual function}

For a QCQP minimization problem, the Lagrangian $\mathcal{L}$ can be written as
\begin{equation}
\label{eq:lagrangian}
    \mathcal{L} (|\mathbf{T}\rangle, \lambda) = 
    \begin{bmatrix}
    \bra{\mathbf{T}} & \bra{\mathbf{S}} 
    \end{bmatrix}
    \begin{bmatrix}
    -Z^{TT}(\lambda) & Z^{TS}(\lambda) \\
    Z^{ST}(\lambda) & Z^{SS}(\lambda)  
    \end{bmatrix} 
    \begin{bmatrix}
    \ket{\mathbf{T}} \\
    \ket{\mathbf{S}} 
    \end{bmatrix},
\end{equation}
where \(\ket{\mathbf{T}}\) is the field over which the primal problem is optimized over (often it is proportional to the polarization current in the design region), \(\ket{\mathbf{S}}\) is some fixed vector of the problem (often proportional to the initial current or electric field), and $\lambda$ is a Lagrange multiplier.
Here, $Z^{TT}$ is Hermitian and $Z^{TS} = Z^{ST\dagger}.$
The stationary point of the Lagrangian for negative definite \(Z^{TT}\) is at \(\ket{\mathbf{T}_{\textrm{opt}}} = Z^{TT-1} Z^{TS}\ket{\mathbf{S}}\), so that the dual function $\mathcal{D}(\lambda)$ and its derivative with respect to the Lagrange multiplier $\lambda$ is given by
\begin{equation} \begin{aligned}
    \mathcal{D}(\lambda) &= \inf_{\ket{\mathbf{T}}} \mathcal{L} (\ket{\mathbf{T}}, \lambda) \\
    \\
    &=\bra{\mathbf{S}} (Z^{ST} Z^{TT-1} Z^{TS} + Z^{SS}) \ket{\mathbf{S}}, \\
    \frac{\partial\mathcal{D}}{\partial \lambda} &= 2 \Re \left( \bra{\mathbf{T}_{\textrm{opt}}} \frac{\partial Z^{TS}}{\partial \lambda} \ket{\mathbf{S}} \right) - \bra{\mathbf{T}_{\textrm{opt}}} \frac{\partial Z^{TT}}{\partial \lambda} \ket{\mathbf{T}_{\textrm{opt}}} + \bra{\mathbf{S}} \frac{\partial Z^{SS}}{\partial \lambda} \ket{\mathbf{S}}.
    \label{eq:dual_ZTT}
\end{aligned} \end{equation}
The best Lagrange dual lower bound follows by computing the supremum of $\mathcal{D}(\lambda)$, i.e., solving the dual problem. Since the dual function $\mathcal{D}(\lambda)$ is concave~\cite{boyd_convex_2004}, the optimal value can be numerically computed using, e.g., gradient descent methods.

In the case of a QCQP maximization problem, $Z^{TT}$ would be positive definite, the dual function would be defined as $\sup_{\ket{\mathbf{T}}}\mathcal{L}(\ket{\mathbf{T}},\lambda)$ and would be a convex function, and the dual problem would be to find the infimum of $\mathcal{D}(\lambda)$ to find the best Lagrange dual upper bound.

\bibliography{refs}